\documentclass[aps,prl,twocolumn,superscriptaddress,nofootinbib]{revtex4-2}
\usepackage[english]{babel}
\usepackage{bm}
\usepackage{amsfonts}
\usepackage{amsmath}
\usepackage{amssymb}
\usepackage{dutchcal}
\usepackage[colorlinks=true, allcolors=cyan]{hyperref}
\usepackage{diagbox}
\usepackage{multirow}
\usepackage{threeparttable}
\usepackage{xfrac}
\usepackage{xcolor}
\usepackage{yfonts}

\begin{document}
\title{Quasi-bound layer-breathing phonons inside perfect dislocations of lattice-relaxed twisted bilayers}
\author{V.~V.~Enaldiev}
\email{vova.enaldiev@gmail.com}
\affiliation{Moscow Center for Advanced Studies, Kulakova str. 20, Moscow 123592, Russia} 
\affiliation{Kotelnikov Institute of Radio-engineering and Electronics of the RAS, Mokhovaya 11-7, Moscow 125009, Russia}

\begin{abstract}
 Using multiscale modelling we investigate layer-breathing phonons in MX$_2$ bilayers (M=Mo,W; X=S,Se) containing dislocations specific for lattice-relaxed moir\'e superlattices. The dislocations, forming in the bilayers with parallel and antiparallel alignment of layers, bring about spatial modulation of the interlayer distance, generating effective potentials that promote the emergence of one-dimensional quasi-bound bands of layer-breathing modes inside perfect dislocations. For parallel MX$_2$ bilayers, perfect dislocations host multiple quasi-bound bands, with frequencies above the layer-breathing mode in rhombohedral-stacked domains. In contrast, antiparallel bilayers exhibit only a single quasi-bound band arising for orientations close to the edge-dislocation type, having frequencies above the layer-breathing mode in 2H-stacked domains.
\end{abstract}

\maketitle

Layer-breathing phonons -- relative out-of-plane oscillations of neighboring layers in van der Waals crystals -- provide important information about details of interlayer coupling. In uniformly stacked bilayers, such as Bernal graphene \cite{Mostaani2015} and hexagonal (2H)/rhombohedral (3R) transition metal dichalcogenide \cite{Enaldiev_PRL} (TMD) bilayers, the frequency of the Brillouin zone center layer-breathing phonon is determined solely by interlayer force parameter, characterized by curvature of interlayer adhesion \cite{Mostaani2015,Enaldiev_PRL} at equilibrium distance, and the reduced mass of layers' unit cells. Although selection rules make the breathing mode Raman active \cite{Saito2016,Lin2017}, the small interlayer force parameters, compared to intralayer ones, place this mode in the far-infrared spectral range, making experimental detection more challenging \cite{Lui2014}. Nevertheless, the zero- momentum layer-breathing modes have been successfully observed in graphene \cite{Lui2013,Lui2014,Lui2015} and TMD \cite{Zhao2013,Chen2015,Baren2019,Boora2024} multilayers using Raman spectroscopy. 

Moir\'e superlattices, forming at the interface of two rigidly rotated layers, generate mesoscale potentials for charge carriers, lattice vibrations etc. reducing spectra to miniband structure. For charge carriers in twisted graphene and TMD bilayers this leads to emergent superconductivity \cite{cao2018unconventional,Yankowitz2019,Xia2024,Guo2025} and strongly correlated states \cite{cao2018correlated,Kennes2021,Foutty2023,Campbell2024,Li2021a} promoted by formation of flat minibands \cite{Bistritzer2011,AitorPRB2021,Naik2020,Magorrian2021,Zhang2020}. 

The spatial variation of stacking arrangements in moir\'e superlattices modifies lattice dynamics in twisted bilayers. For short-period moir\'e patterns, specific for large interlayer twist angles and/or layers' lattice parameter mismatches, miniband folding of phonon spectra can be described in a rigid lattice approximation \cite{BalandinPRB2013,Song2019}, whereas for long-period moir\'e superlattices relaxation of atomic positions in the layers should also be taken into account \cite{KoshinoPRB2019,Quan2021,Maity2022}. 

For twisted TMD bilayers, the relaxed moir\'e patterns differ for parallel (P) and antiparallel (AP) alignments of the constituent monolayers. For small-angle twisted P bilayers the moir\'e superlattice relaxes into an array of triangular 3R-stacked domains \cite{Weston2020,rosenberger2020}, separated by network of partial dislocations \cite{Enaldiev_PRL,CarrPRB2018,Engelke2023}. The ferroelectric polarization of these domains \cite{Weston2022,Ferreira2021} enables manipulation of the dislocation network, including the merging of partial dislocation pairs into perfect dislocations \cite{Enaldiev2022,Ko2023,Molino2023}. In contrast, twisted AP bilayers with small twist angles form relaxed moir\'e pattern consisting of hexagonal 2H-stacked domains \cite{Weston2020,Enaldiev_PRL}, divided by a network of perfect dislocations.   

Phonons in twisted P and AP bilayers, have been extensively studied both theoretically \cite{KoshinoPRB2019,PRB2019Ochoa,Maity2020,PRB2022_Samajdar,Liu2022,Jong2022,Lu2022,PRB2023Cappelluti,PRB2023Girotto} and experimentally \cite{Quan2021,Jong2022}. These studies  usually assume the translational and $C_3$-rotation symmetries of ideal moir\'e superlattices that only permit screw-type dislocations in relaxed moir\'e patterns \cite{Enaldiev_PRL}. However, in experimentally investigated samples with long-period moir\'e superlattices (see e.g. \cite{Butz2013,Weston2020,Molino2023,Liang2023,Engelke2023}) both of the symmetries are often broken, resulting in formation of mixed and edge perfect/partial dislocations in the networks. 

In the work, we focus on individual dislocations in these networks and demonstrate that layer-breathing phonons can form quasi-bound one-dimensional (1D) bands inside perfect dislocations for both P and AP TMD bilayers. Unlike in-plane interdomains phonons \cite{Enaldiev2025}, which emerge in the gap of bulk shear modes, these quasi-bound modes arise in the 2D continuum of layer-breathing phonons due to an effective double-barrier potential created by perfect dislocations. 

Using MoS$_2$ as a reference material, we find that perfect dislocations in P bilayers give rise to multiple quasi-bound 1D bands of layer-breathing phonons,  whereas AP bilayers host only a single quasi-bound band near the edge dislocation orientation. Furthermore, while individual partial dislocations cannot confine layer-breathing phonons, aligned partial dislocation pairs in close proximity can generate double-barrier potentials similar to that of perfect dislocations in P bilayers, thereby enabling the formation of quasi-bound bands.  

%%%%%%%%%%%%%%%%%%%%%%%%%%%%%%%%%%%%%%%%
%%%%%%%%%%%%%%%%%%%%%%%%%%%%%%%%%%%%%%%%
\begin{figure}%[!th]
	%\centering
	\includegraphics[width=1.0\columnwidth]{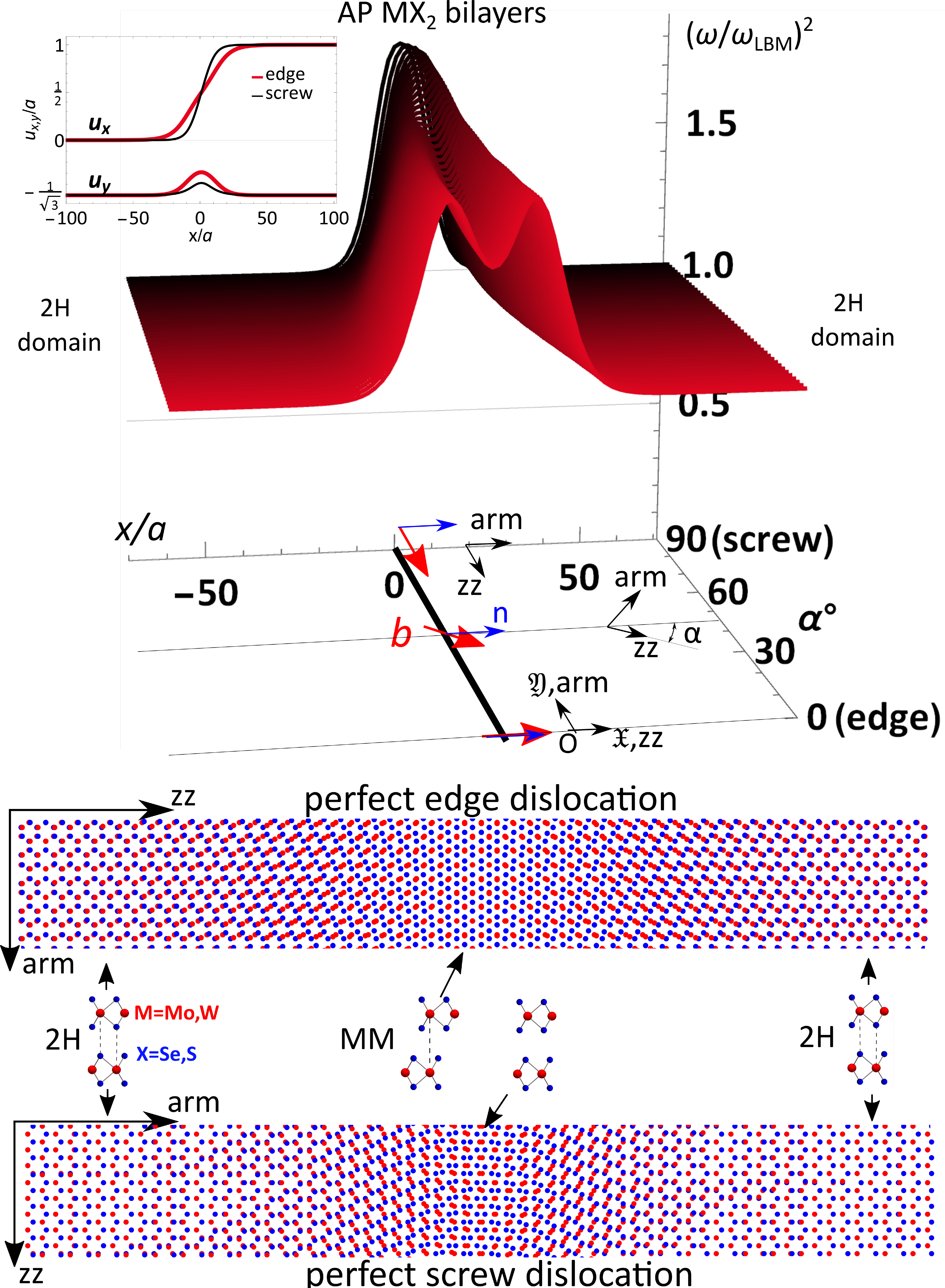}
	\caption{Top panel represents effective potentials for layer-breathing phonons, formed by perfect dislocations (from edge (red) to screw (black) types) separating 2H stacking domains, in AP MX$_2$ bilayers . Armchair (arm) and zigzag (zz) crystallographic axes are shown by black arrows; Burgers vector $|\bm{b}|=a$ (aligned with zigzag) and normal to dislocation axis $\bm{n}$ are highlighted by red and blue arrows, respectively. Inset shows distributions of the in-plane displacement fields across the edge and screw perfect dislocations in AP bilayers. Bottom panel shows modulations of atomic structures across perfect edge and screw dislocations.\label{fig:Potential}}
\end{figure}
%%%%%%%%%%%%%%%%%%%%%%%%%%%%%%%%%%%%%%%
%%%%%%%%%%%%%%%%%%%%%%%%%%%%%%%%%%%%%%%

To study layer-breathing phonons in MX$_2$ bilayers containing dislocations we use a multiscale approach combining elasticity theory with an interpolation function for interlayer adhesion energy. This relies on the following Lagrange function: 
\begin{align}\label{Eq:1}
	\mathcal{L}&= \dfrac{\rho\dot{z}^{2}}{4}\\
	& - \dfrac{\kappa}{4}\left[(\Delta z)^2+2(1-\sigma)\left\{(\partial^2_{xx}z)(\partial^2_{yy}z)-(\partial^2_{xy}z)^2\right\} \right] \nonumber\\ 
	&-\dfrac{\lambda}{4}\left(u_{ii}\right)^2 - \dfrac{\mu}{2}\left(u_{ij}\right)^2
	 - W_{\rm ad}(\bm{u},d(\bm{u})+z). \nonumber 
\end{align}  
On the right-hand side of Eq. \eqref{Eq:1} the first line represents the kinetic energy of relative out-of-plane displacements, $z(\bm{r},t)$, of the layers characterized by monolayer's areal reduced mass density $\rho/2$; the second line corresponds to the bending energy of the displacements, determined by the layers' bending rigidity $\kappa$ and Poisson ratio $\sigma$; $\Delta=\partial^2_{xx}+\partial^2_{yy}$ is 2D Laplace operator for in-plane coordinates $\bm{r}=(x,y)$. The first two terms in the last line describe in-plane elastic energy, determined by layers' Lam\'e parameters $\lambda$, $\mu$ and strain $2u_{ij}=\partial u_i/\partial x_j+\partial u_j/\partial x_i$ ($i,j=x,y$), produced by the dislocation displacement field $\bm{u}(\bm{r})=(u_x,u_y)$. The final term represents interlayer adhesion energy density, $W_{\rm ad}$, that depends on the stacking configuration set by $\bm{u}$ and the variation of interlayer distance, $d(\bm{u})+z$, across the dislocation \cite{Zhou2015,Enaldiev_PRL}. Following Ref. \cite{Enaldiev_PRL}, explicit expression for $W_{\rm ad}$ of MX$_2$ bilayers reads as follows: 
\begin{multline}\label{Eq:2}
	 W_{\rm ad}(\bm{r}_0,d=d_0+\delta) = f(d_0+\delta) +\\
	  \sum_{l=1,2,3}\left\{w_1e^{-q\delta}\cos(\bm{G}_l\bm{r}_0)+w_2e^{-G\delta}\sin(\bm{G}_l\bm{r}_0+\varphi_{\rm P/AP})\right\}\approx\\
	f(d_0)+\varepsilon\delta^2 +
	  \sum_{l=1,2,3}\left\{w_1\left[1-q\delta+\frac{(q\delta)^2}{2}\right]\cos(\bm{G}_l\bm{r}_0)\right.\\
	  \left.+w_2\left[1-G\delta+\frac{(G\delta)^2}{2}\right]\sin(\bm{G}_l\bm{r}_0+\varphi_{\rm P/AP})\right\}.
\end{multline}
Here, $\bm{r}_0$ is in-plane interlayer offset, distance $d=d_0+\delta$ between the layers is counted with respect to optimal interlayer distance, $d_0$, of stacking-averaged adhesion energy, $f(d)\approx f(d_0)+\varepsilon \delta^2$, $\bm{G}_{1}=G(\sqrt{3}/2,1/2)$, $\bm{G}_{2,3}=\hat{R}^{\mp1}_{3}\bm{G}_{1}$ are triad of the shortest reciprocal vectors of bilayers ($G=\frac{4\pi}{a\sqrt{3}}$, $\hat{R}^{-1}_{3}$ ($\hat{R}^{1}_{3}$) is clockwise (anticlockwise) 120$^{\circ}$-rotation, $a$ is the monolayer lattice parameter), the phase $\varphi_{\rm AP}=0$ and $\varphi_{\rm P}=\pi/2$ takes into account symmetry of the adhesion energy for P and AP alignments of the layers.  $w_{1,2}$, $q$, $d_0$, and $\varepsilon$ are material-specific parameters whose values can be found in Ref. \cite{Enaldiev_PRL}. Unlike the linear expansion in $\delta$ of the stacking-dependent terms used in Ref. \cite{Enaldiev_PRL} to study relaxation of moir\'e pattern in twisted MX$_2$ bilayers, after the second equality in Eq. \eqref{Eq:2} we also take into account $\delta^2$ terms necessary for the harmonic approximation on the relative out-of-plane deformations of the layers.  

To establish the relationship between lateral offsets and interlayer distances in Eq. \eqref{Eq:2}, we employ an optimal interlayer distance approximation \cite{Zhou2015,CarrPRB2018,Enaldiev_PRL}, which reduces to minimization of the $W_{\rm ad}(\bm{r}_0,d_0+\delta)$ \eqref{Eq:2} with respect to $\delta$. This results in 
\begin{align}
	\delta(\bm{r}_0)&=\sum_{\substack{l=1,2,3}}\frac{\left[w_1q\cos(\bm{G}_l\bm{r}_0)+w_2G\sin(\bm{G}_l\bm{r}_0+\varphi_{\rm P/AP})\right]}{V(\bm{r}_0)}, \label{Eq:3}
\end{align} 
where
\begin{multline}\label{Eq:potential_gen}
	V(\bm{r}_0)=2\varepsilon + \\ \sum_{l=1,2,3}\left[w_1q^2\cos(\bm{G}_l\bm{r}_0)+w_2G^2\sin(\bm{G}_l\bm{r}_0+\varphi_{\rm P/AP})\right].
\end{multline}
 
To analyze layer-breathing phonons in the bilayers with an individual dislocation we substitute Eqs. \eqref{Eq:2}-\eqref{Eq:potential_gen} into $\mathcal{L}$ \eqref{Eq:1} and transform the adhesion energy to the local stacking via replacements $\bm{r}_0\to\bm{u}$ and $\delta(\bm{r}_0)\to\delta(\bm{u})+z(\bm{r},t)$. 

We begin by analyzing perfect dislocations in AP MX$_2$ bilayers. For a fixed Cartesian coordinate system $\textswab{X}O\textswab{Y}$, where $O\textswab{X}$ ($O\textswab{Y}$) aligns with the zigzag (armchair) crystal axis, we define a Burgers vector $\bm{b}=(a,0)$ and a unit vector $\bm{n}_{\alpha}=(\cos\alpha,\sin\alpha)$ perpendicular to the dislocation axis. Here, $\alpha^{\circ}=0^{\circ}$ corresponds to an edge dislocation while $\alpha^{\circ}=90^{\circ}$ represents a screw dislocation (see Fig. \ref{fig:Potential}). 

For each angle $\alpha$ we establish a local Cartesian system $xOy$, with $Ox$ parallel to $\bm{n}_{\alpha}$. The displacement field $\bm{u}=\bm{u}^{(\alpha)}(x)$ results from minimization of the Lagrangian $L=\int\mathcal{L}d^2\bm{r}dt$ with $z(\bm{r},t)\equiv 0$ and boundary conditions $\bm{u}^{(\alpha)}(x=\pm\infty)=(0,-a/\sqrt{3})$ that enforce 2H-stacking in domains characteristic for AP MX$_2$ bilayers \cite{Weston2020}. This displacement fields have been previously determined in Refs. \cite{Enaldiev_PRL, Enaldiev2024}, with modulation the atomic structures across the dislocations shown on bottom panel in Fig. \ref{fig:Potential}. We therefore treat dislocation displacement fields as known vector functions.

Including the out-of-plane displacement $z(\bm{r},t)$ in the Lagrangian $L$, leads to the following Lagrange-Euler equation:
\begin{equation}\label{Eq:4}
\ddot{z}+\frac{\kappa}{\rho}\Delta^2z+\frac{2}{\rho}{V\rm }\left(\bm{u}^{(\alpha)}(x)\right)z=0,
\end{equation}
where $V$ is determined by Eq. \eqref{Eq:potential_gen} for local stacking set by the displacement field $\bm{u}^{(\alpha)}(x)$. As it follows from Eq. \eqref{Eq:4}, a perfect dislocation forms an effective orientation-dependent potential for layer-breathing modes,  shown in Fig. \ref{fig:Potential}. The spatial modulation of $V$ arises specifically from the $\delta^2$ terms in Eq. \eqref{Eq:2}. Inside domains ($x=\pm\infty$), Eq. \eqref{Eq:4} describes continuum of layer-breathing phonons characterized by the spectral gap $\omega_{\rm LBM}=\sqrt{(4\varepsilon-3w_1q^2-3\sqrt{3}w_2G^2)/\rho}$. 

While screw dislocations create a simple potential barrier causing off-resonant scattering, edge dislocations develop a local minimum at the core that, as will be shown below, supports a 1D band of quasi-bound modes propagating along the dislocation line. This orientation dependence stems from the enhanced tangential displacement component in the core region (see inset on top panel in Fig. \ref{fig:Potential}).

%%%%%%%%%%%%%%%%%%%%%%%%%%%%%%%%%%%%%%%%
%%%%%%%%%%%%%%%%%%%%%%%%%%%%%%%%%%%%%%%%
\begin{figure}%[!th]
	%\centering
	\includegraphics[width=1.0\columnwidth]{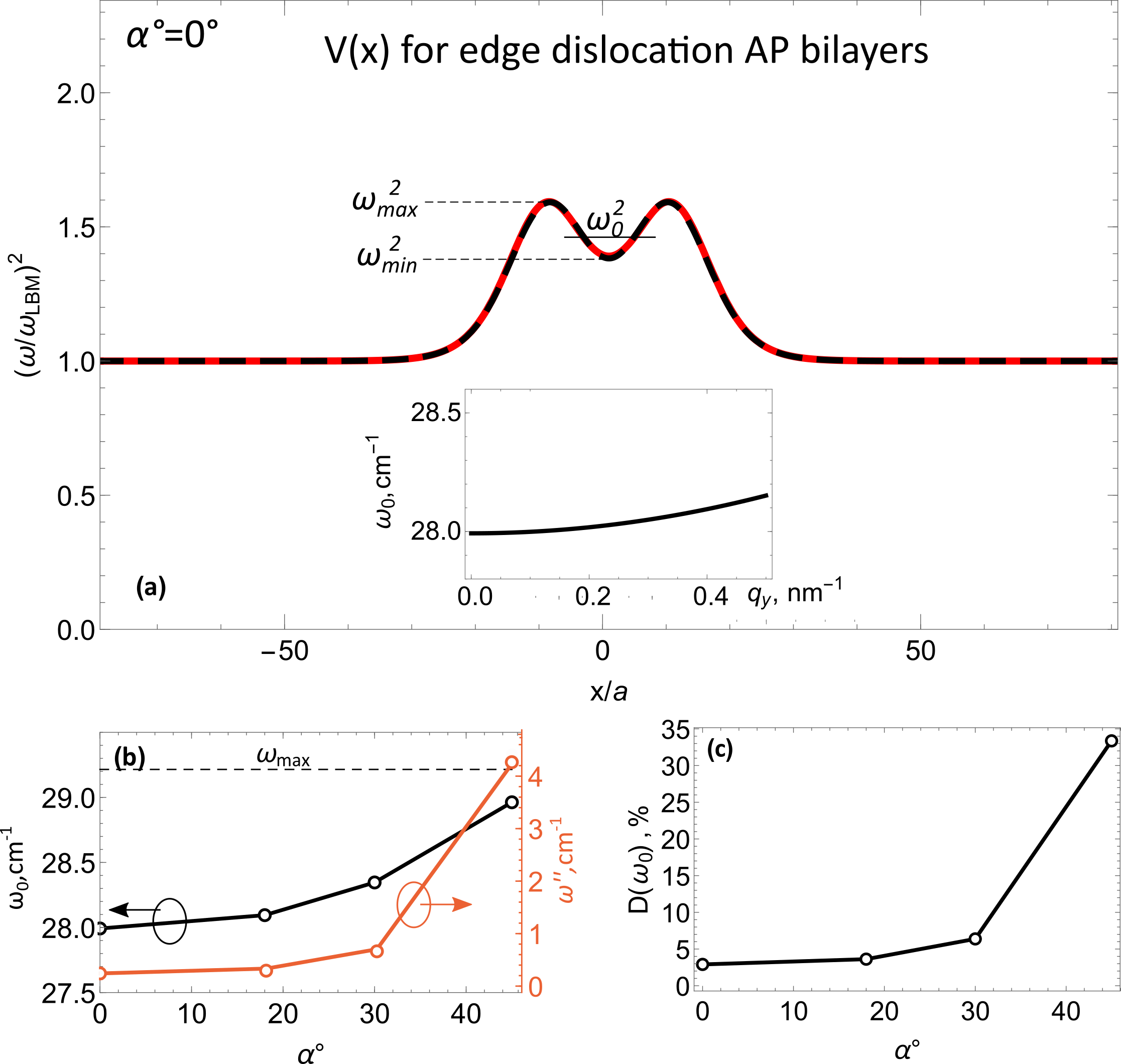}
	\caption{(a) Red solid (black dashed) line demonstrates the exact (fitted \eqref{Eq:potential_edge_d}) effective potential produced by the edge dislocation in AP MoS$_2$ bilayers for layer-breathing phonons. (b) Dependences of the 1D band bottom frequency, decay rate and (c) transparency of the barrier (at $\omega_0$) \eqref{Eq:transparency} on the perfect dislocation orientation with respect to Burgers vector in AP MoS$_2$.  \label{fig:spectrum}}
\end{figure}
%%%%%%%%%%%%%%%%%%%%%%%%%%%%%%%%%%%%%%%
%%%%%%%%%%%%%%%%%%%%%%%%%%%%%%%%%%%%%%%

To demonstrate formation of the 1D quasi-bound band, we consider the edge dislocation ($\alpha^{\circ}=0^{\circ}$), for which the effective potential can be fitted as follows (see Fig. \ref{fig:spectrum}(a)):
\begin{equation}\label{Eq:potential_edge_d}
	\frac{2}{\rho}V\left(\bm{u}^{(0)}(x)\right)=\omega^2_{\rm LBM}+\frac{A_1}{1+B_1\cosh\left(\frac{x}{l_1}\right)}-A_2e^{-\left(\frac{x}{l_2}\right)^2},
\end{equation}
where $A_{1,2}$, $B_1$, and $l_{1,2}$ are material-dependent fitting parameters. Substituting in Eq. \eqref{Eq:4} parameters for AP MoS$_2$ from Ref. \cite{Enaldiev_PRL}, one finds $\omega_{\rm LBM} \approx 23$\, cm$^{-1}$, while the fitted parameters are $l_1\approx 1.17$\,nm, $l_2\approx2.91$\,nm,  $A_1\approx 581$\,cm$^{-2}$, $A_2\approx 349$\,cm$^{-2}$, $B_1=0.05$. 

We look for solution of Eqs. \eqref{Eq:4}, \eqref{Eq:potential_edge_d} in the form $z(\bm{r},t)=\xi(x)e^{iq_{y}y-i\omega t}$, where $q_{y}$ is a conserving tangential wavenumber and complex frequency $\omega=\omega'+i\omega''$, with $\omega'$ and  $\omega''$ characterizing mode frequency and decay rate into continuum, respectively. To determine $\omega'$ we use a parabolic approximation for the effective potential in vicinity of the dislocation core ($x=0$), $2V/\rho\approx\omega^2_{\rm min}+A_2(x/l_2)^2$ ($\omega^2_{\rm min}\equiv\omega^2_{\rm LBM}+A_1/(1+B_1)-A_2$). 
Applying Fourier transform to Eq. \eqref{Eq:4} one obtains 1D Schroedinger equation in momentum space for $\widetilde{\xi}(q_x)=\int\xi(x)e^{-iq_xx}dx$:
\begin{equation}\label{Eq:Fourier}
	-\frac{A_2}{l_2^2}\frac{d^2}{dq_x^2}\widetilde{\xi}+\frac{1}{(2\pi c)^2}\frac{\kappa}{\rho}(q_{x}^2+q_y^2)^2\widetilde{\xi}=\left(\omega'^2-\omega_{\rm min}^2\right)\widetilde{\xi}
\end{equation}   
where factor $(2\pi c)^{-2}$ in front of the second term appears due to units (cm$^{-2}$) of the potential. Eigen modes of Eq. \eqref{Eq:Fourier} are either even or odd with respect to $q_x\to-q_x$. Putting $q_y=0$, we find eigenvalues of Eq. \eqref{Eq:Fourier} using shooting method with vanishing boundary condition at $q_x=+\infty$, relevant for the parabolic approximation. For AP MoS$_2$ bilayers, this results in emergence of the only ground mode $\omega'=\omega_0\approx 28$\,cm$^{-1}$ lower the effective potential maximum, $\omega^2_{\rm max}\approx(29.22)^2$ cm$^{-2}$. Taking into account $q_y$-dependent terms perturbatively, we obtain the following 1D band dispersion, 
\begin{equation}\label{Eq:bandspectra}
	\omega'_{q_y}=\sqrt{\omega_0^2+\frac{\kappa q_y^2\left(2\langle q_x^2\rangle+q_y^2\right)}{\rho(2\pi c)^2}},%\approx\omega_0+\kappa q_y^2\langle q_x^2\rangle/(2\pi c)^2\rho\omega_0
\end{equation}
depicted on the inset in Fig. \ref{fig:spectrum}(a). 

%%%%%%%%%%%%%%%%%%%%%%%%%%%%%%%%%%%%%%%%
%%%%%%%%%%%%%%%%%%%%%%%%%%%%%%%%%%%%%%%%
\begin{figure*}%[!th]
	%\centering
	\includegraphics[width=2.0\columnwidth]{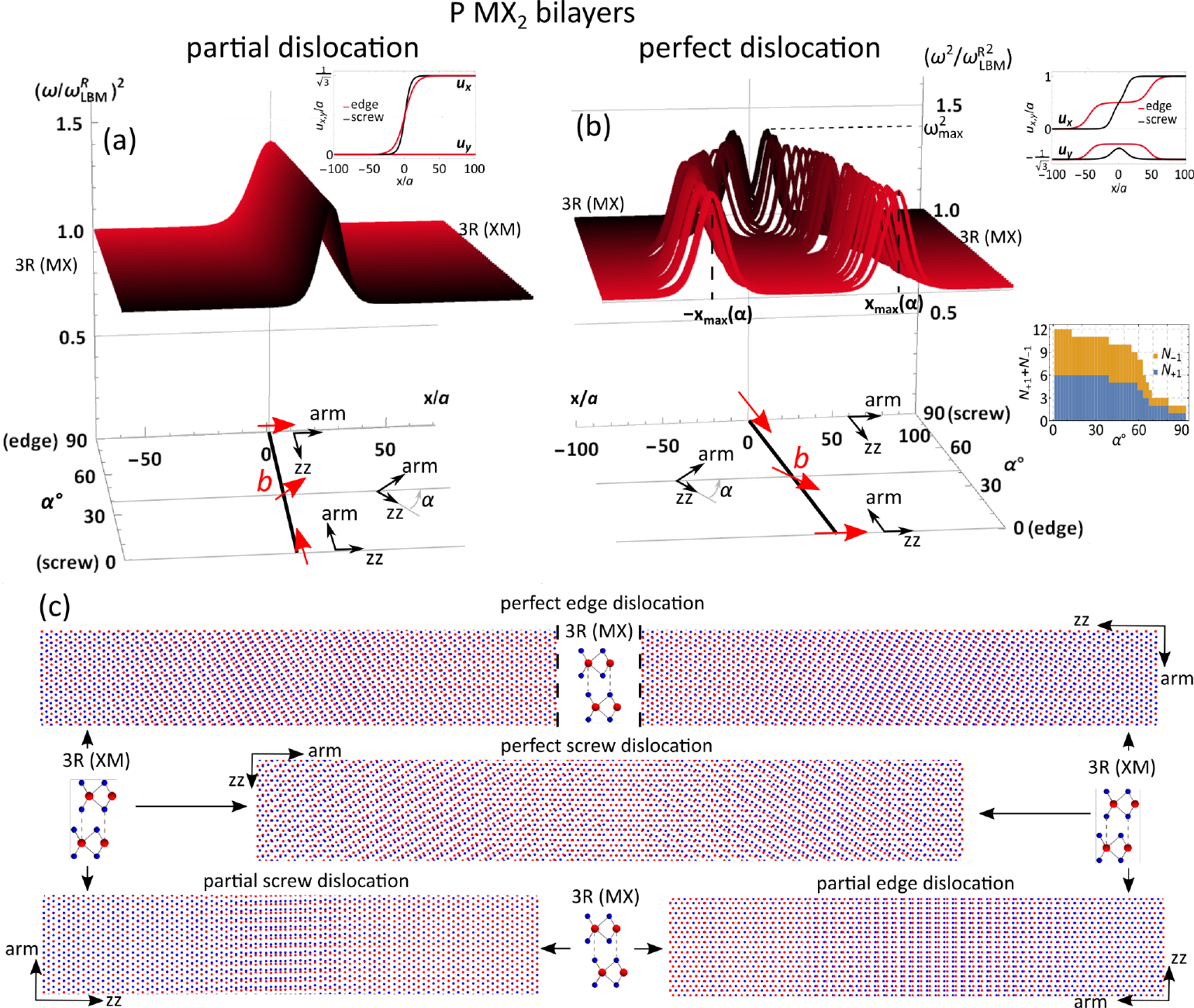}
	\caption{Effective potentials for layer-breathing phonons, formed by partial ($|\bm{b}|=a/\sqrt{3}$) (a) and perfect ($|\bm{b}|=a$) (b) dislocations (from screw (black) to edge (red) types), in P MX$_2$ bilayers. Top insets  demonstrate displacement fields across screw and edge dislocations. Bottom inset in (b) displays dependence of the number of even, $N_{+1}$, and odd, $N_{-1}$, 1D quasi-bound bands formed at perfect dislocations in P MX$_2$ bilayers. (c) Modulation of atomic structures across perfect/partial screw and edge dislocations in P TMD bilayers.  \label{fig:VPbilayers}}
\end{figure*}
%%%%%%%%%%%%%%%%%%%%%%%%%%%%%%%%%%%%%%%
%%%%%%%%%%%%%%%%%%%%%%%%%%%%%%%%%%%%%%%

Next, we estimate decay rates of the 1D modes, demonstrating their well-defined nature not only for pure edge dislocations but also for orientations within $|\alpha^{\circ}|\lesssim 30^{\circ}$. Decay of the 1D modes into continuum results from finite transparency of the barrier between the local minimum and value of $V$ in domains (see Fig. \ref{fig:spectrum}(a)). We estimate the decay rate as follows,
\begin{equation}\label{Eq:decayrate}
	\omega''=2\sqrt{\left(\omega'^2-\omega_{\rm min}^2\right)}D(\omega').
\end{equation} 
Here, $D(\omega')$ is probability to tunnel through the left/right barrier for the mode with frequency $\omega'$, and the prefactor $2\sqrt{\left(\omega'^2-\omega_{\rm min}^2\right)}$ corresponds to the collision frequency with barriers. Using analogy with $\alpha$-decay theory \cite{gamov1928} we calculate $D$ via an underbarrier motion integral for the parabolic approximation of the barrier $2V/\rho\approx\omega_{\rm max}^2-\eta(x'-x_{\rm max})^2$ around its maximum $\omega_{\rm max}^2$. This leads to
\begin{eqnarray}\label{Eq:transparency}
	D(\omega')\approx\qquad\qquad\qquad\qquad\qquad\qquad\qquad\qquad\qquad\qquad\quad\\ 
	\left|\exp\left(-e^{i\frac{\pi}{4}}\sqrt{2\pi c}\int_{-\Delta x}^{+\Delta x}\left[\frac{\rho}{\kappa}\left(\omega_{\rm max}^2-\eta x'^2-\omega'^2\right)\right]^{1/4}dx'\right)\right|^2\nonumber \\ 
	=\exp\left(-\frac{2\pi}{3}\frac{\Gamma(1/4)}{\Gamma(3/4)}\left(\frac{c^2\rho}{\kappa}\right)^{1/4}\frac{\left(\omega_{\rm max}^2-\omega'^2\right)^{3/4}}{\sqrt{\eta}}\right), \nonumber
\end{eqnarray}
where $\Delta x=\left(\omega^2_{\rm max}-\omega'^2\right)^{1/2}/\sqrt{\eta}$ and $\Gamma(x)$ is the Gamma function. The decay rates and barrier transparencies of the $\omega_0$-mode, depicted in Fig. \ref{fig:spectrum}(b) and \ref{fig:spectrum}(c) as functions of perfect dislocation orientation, demonstrate formation of long-living quasi-bound 1D modes in the interval $|\alpha^{\circ}|\lesssim 30^{\circ}$ around the edge orientation for AP MoS$_2$.

Second, we consider P MX$_2$ bilayers, where partial/perfect dislocations separate domains with rhombohedral stacking arrangements of the layers \cite{Weston2020,rosenberger2020}, illustrated in Fig. \ref{fig:VPbilayers}(c). The displacement fields of individual partial dislocations, characterized by Burgers vector $\bm{b}=(0,a/\sqrt{3})$ in the fixed Cartesian system, are described by only tangential components regardless dislocation type \cite{Enaldiev2024} (see inset in Fig. \ref{fig:VPbilayers}(a)). Consequently, the resulting effective potentials form simple barriers, depicted in Fig. \ref{fig:VPbilayers}(a), that support only off-resonant scattering modes of layer-breathing phonons. However, applied strain can induce stripe-patterned moir\'e superlattices with aligned partial dislocation arrays \cite{Butz2013}. In such configurations, neighboring partial dislocation pairs may host 1D quasi-bound bands analogous to the AP case.

Under external stimuli, for instance out-of-plane electric field \cite{Enaldiev2022,Molino2023,Ko2023}, pairs of partial dislocations in P bilayers can also merge into perfect dislocations, whose atomic structures for edge and screw orientation are shown in Fig. \ref{fig:VPbilayers}(c). As in AP case, the perfect dislocations are characterized by both tangential and normal components of the displacement fields, that give rise to double-barrier effective potentials for layer-breathing phonons, shown in Fig. \ref{fig:VPbilayers}(b). For each dislocation orientation the potential enables formation of 1D quasi-bound bands between the barriers. Not interested in finding exact eigen frequencies given by Eq. \eqref{Eq:4}, we estimate the number of arising 1D quasi-bound bands, approximating the potentials by rectangular 1D wells with bottoms equal to $V(\bm{u}^{(\alpha)}(0))\equiv V(0)$ and impenetrable walls at the barrier maxima, $\pm x_{\rm max}(\alpha)$ (see Fig. \ref{fig:VPbilayers}(b)). Using parity of the potentials, $V(x)=V(-x)$, we obtain the following quantization rules:
\begin{align}
	\omega'_{\nu,n}(q_y)&=\sqrt{V(0)+\frac{\kappa}{\rho(2\pi c)^2}\left(q_y^2+\frac{\pi^2(n-\frac{1}{2}\delta_{\nu,1})^2}{x_{\rm max}^2(\alpha)}\right)^2}, 
\end{align}  
where parity eigen value $\nu=1$ ($\nu=-1$) for even (odd) bands, $n=1,...,N_{\nu}(\alpha)$ are quantization numbers, with $N_{\nu}(\alpha)$ is the largest satisfying $\omega'_{\nu,N_{\nu}(\alpha)}(0)< \omega_{\rm max}$; $\delta_{\nu,1}$ is Kronecker symbol. We find that the number of 1D quasi-bound bands emerging inside perfect dislocations correlates with their widths: multiple even and odd quasi-bound bands emerge for the widest edge dislocations, whereas only one of each are formed for the screw dislocations (see inset in Fig. \ref{fig:VPbilayers}(b)). 

Formation of the 1D modes inside perfect dislocations leads to increase of average interlayer distance due to fluctuations. To estimate the effect we consider a single 1D band inside a perfect edge dislocation in AP bilayers, described by the long-wavelength spectrum \eqref{Eq:bandspectra}, $\omega_{q_y}\approx \omega_0+\kappa\langle q_x^2\rangle q_y^2/\rho \omega_0(2\pi c)^2$. For a dislocation of length $\ell$, we quantize the modes as follows:
\begin{equation}\label{Eq:1Dquants}
	z_{1D}(y,t)=\sum_{q_y}\sqrt{\frac{\hbar}{\rho \zeta \ell\omega_{q_y}}}\left(a^{+}_{q_y}e^{i\omega_{q_y}t}+a_{q_y}e^{-i\omega_{q_y}t}\right)e^{iq_yy},
\end{equation}
where $\zeta$ is transversal decay length, $a_{q_y}$ and $a^{+}_{q_y}$ are destruction and creation operators of the 1D modes. The thermally averaged over squared displacement \eqref{Eq:1Dquants} reads
\begin{multline}\label{Eq:1Dfluc}
	\langle z_{1D}^2\rangle_{T}=\frac{2\hbar c}{\zeta\sqrt{\rho\kappa\langle q_x^2\rangle}}\left[\int_{1}^{x_*}\frac{dx}{x\sqrt{x-1}}\frac{1}{\left(e^{\frac{\hbar\omega_0}{T}x}-1\right)}+\right.\\
\left.	+\arccos\left(\frac{1}{\sqrt{x_*}}\right)\right],
\end{multline}
where $x_{*}=1+\kappa\langle q_x^2\rangle (G/2)^2/\rho \omega_0^2(2\pi c)^2$ is ratio of zero and edge momenta ($G/2$) frequencies for the 1D band, and $T$ is the lattice temperature. 

\begin{figure}%[!th]
	%\centering
	\includegraphics[width=1.0\columnwidth]{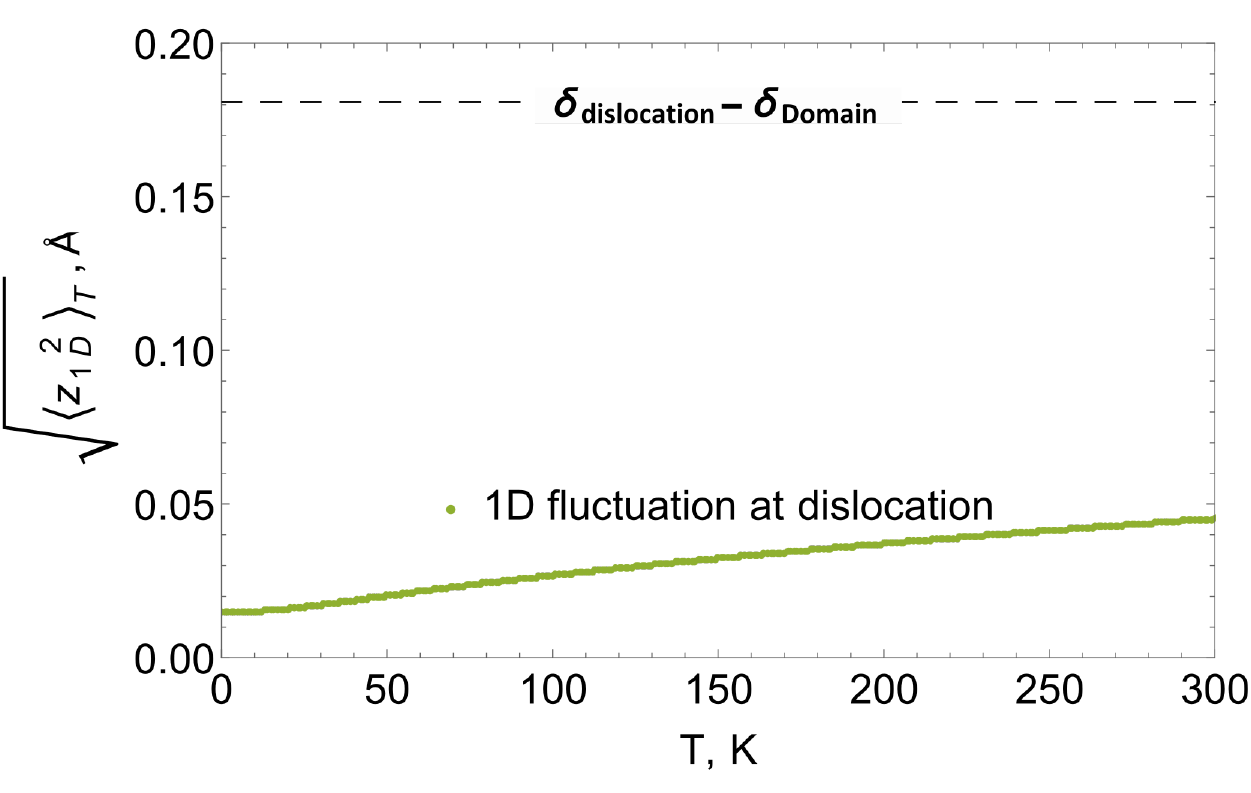}
	\caption{Temperature dependence of interlayer distance fluctuation at the edge dislocation in AP MoS$_2$. Dashed line indicates variation  of optimal interlayer distance due to interlayer adhesion, Eq. \eqref{Eq:3}. \label{fig:fluct}}
\end{figure}

As shown in Fig. \ref{fig:fluct}, these fluctuations induce only a minor increase in the interlayer distance at the dislocation core (relative to the adhesion-driven variation, $\delta(\bm{u}(0))-\delta(\bm{u}(-\infty))$) even at room temperature. This validates the optimal interlayer distance approximation used in Eq. \eqref{Eq:4}.

%%%%%%%%%%%%%%%%%%%%%%%%%%%%%%%%%%%%%%%%
%%%%%%%%%%%%%%%%%%%%%%%%%%%%%%%%%%%%%%%%
%%%%%%%%%%%%%%%%%%%%%%%%%%%%%%%%%%%%%%%
%%%%%%%%%%%%%%%%%%%%%%%%%%%%%%%%%%%%%%%

To conclude, we have demonstrated that perfect dislocations in both P and AP TMD bilayers create double-barrier potentials for layer-breathing phonons, resulting in the formation of long-lived 1D quasi-bound bands inside them. These bands exhibit frequencies above those of the layer-breathing modes in their respective rhombohedral (P) and hexagonal (AP) domains. Moreover, orientation-dependent characteristics of these 1D bands enable strain-engineering approaches \cite{Hou2024} for precise spectral tuning. 

While individual partial dislocations in P TMD bilayers cannot support such 1D quasi-bound bands, distorted relaxed moir\'e superlattices with closely aligned partial dislocation pairs, e.g. by external stress, can mimic the effective potentials of perfect dislocations, thereby generating 1D bands within the continuum of 2D layer-breathing phonons. 

Finally, we note that the obtained results are not limited by MX$_2$ bilayers, but are qualitatively valid for graphene and P and AP hBN bilayers as well, as these materials possess similar structure of interlayer adhesion energy \cite{Mostaani2015,Zhou2015,Szyniszewski2025}. 

{\it Acknowledgements.} I acknowledge fruitful discussions with Zaur Alisultanov. The work was supported by the Russian Science Foundation (project No. 24-72-10015).

\bibliography{refer}

\begin{thebibliography}{55}%
\makeatletter
\providecommand \@ifxundefined [1]{%
 \@ifx{#1\undefined}
}%
\providecommand \@ifnum [1]{%
 \ifnum #1\expandafter \@firstoftwo
 \else \expandafter \@secondoftwo
 \fi
}%
\providecommand \@ifx [1]{%
 \ifx #1\expandafter \@firstoftwo
 \else \expandafter \@secondoftwo
 \fi
}%
\providecommand \natexlab [1]{#1}%
\providecommand \enquote  [1]{``#1''}%
\providecommand \bibnamefont  [1]{#1}%
\providecommand \bibfnamefont [1]{#1}%
\providecommand \citenamefont [1]{#1}%
\providecommand \href@noop [0]{\@secondoftwo}%
\providecommand \href [0]{\begingroup \@sanitize@url \@href}%
\providecommand \@href[1]{\@@startlink{#1}\@@href}%
\providecommand \@@href[1]{\endgroup#1\@@endlink}%
\providecommand \@sanitize@url [0]{\catcode `\\12\catcode `\$12\catcode
  `\&12\catcode `\#12\catcode `\^12\catcode `\_12\catcode `\%12\relax}%
\providecommand \@@startlink[1]{}%
\providecommand \@@endlink[0]{}%
\providecommand \url  [0]{\begingroup\@sanitize@url \@url }%
\providecommand \@url [1]{\endgroup\@href {#1}{\urlprefix }}%
\providecommand \urlprefix  [0]{URL }%
\providecommand \Eprint [0]{\href }%
\providecommand \doibase [0]{https://doi.org/}%
\providecommand \selectlanguage [0]{\@gobble}%
\providecommand \bibinfo  [0]{\@secondoftwo}%
\providecommand \bibfield  [0]{\@secondoftwo}%
\providecommand \translation [1]{[#1]}%
\providecommand \BibitemOpen [0]{}%
\providecommand \bibitemStop [0]{}%
\providecommand \bibitemNoStop [0]{.\EOS\space}%
\providecommand \EOS [0]{\spacefactor3000\relax}%
\providecommand \BibitemShut  [1]{\csname bibitem#1\endcsname}%
\let\auto@bib@innerbib\@empty
%</preamble>
\bibitem [{\citenamefont {Mostaani}\ \emph {et~al.}(2015)\citenamefont
  {Mostaani}, \citenamefont {Drummond},\ and\ \citenamefont
  {Fal’ko}}]{Mostaani2015}%
  \BibitemOpen
  \bibfield  {author} {\bibinfo {author} {\bibfnamefont {E.}~\bibnamefont
  {Mostaani}}, \bibinfo {author} {\bibfnamefont {N.}~\bibnamefont {Drummond}},\
  and\ \bibinfo {author} {\bibfnamefont {V.}~\bibnamefont {Fal’ko}},\
  }\bibfield  {title} {\bibinfo {title} {Quantum monte carlo calculation of
  the binding energy of bilayer graphene},\ }\href
  {https://doi.org/10.1103/physrevlett.115.115501} {\bibfield  {journal}
  {\bibinfo  {journal} {Physical Review Letters}\ }\textbf {\bibinfo {volume}
  {115}},\ \bibinfo {pages} {115501} (\bibinfo {year} {2015})}\BibitemShut
  {NoStop}%
\bibitem [{\citenamefont {Enaldiev}\ \emph {et~al.}(2020)\citenamefont
  {Enaldiev}, \citenamefont {Z\'olyomi}, \citenamefont {Yelgel}, \citenamefont
  {Magorrian},\ and\ \citenamefont {Fal'ko}}]{Enaldiev_PRL}%
  \BibitemOpen
  \bibfield  {author} {\bibinfo {author} {\bibfnamefont {V.~V.}\ \bibnamefont
  {Enaldiev}}, \bibinfo {author} {\bibfnamefont {V.}~\bibnamefont {Z\'olyomi}},
  \bibinfo {author} {\bibfnamefont {C.}~\bibnamefont {Yelgel}}, \bibinfo
  {author} {\bibfnamefont {S.~J.}\ \bibnamefont {Magorrian}},\ and\ \bibinfo
  {author} {\bibfnamefont {V.~I.}\ \bibnamefont {Fal'ko}},\ }\bibfield  {title}
  {\bibinfo {title} {Stacking domains and dislocation networks in marginally
  twisted bilayers of transition metal dichalcogenides},\ }\href
  {https://doi.org/10.1103/PhysRevLett.124.206101} {\bibfield  {journal}
  {\bibinfo  {journal} {Phys. Rev. Lett.}\ }\textbf {\bibinfo {volume} {124}},\
  \bibinfo {pages} {206101} (\bibinfo {year} {2020})}\BibitemShut {NoStop}%
\bibitem [{\citenamefont {Saito}\ \emph {et~al.}(2016)\citenamefont {Saito},
  \citenamefont {Tatsumi}, \citenamefont {Huang}, \citenamefont {Ling},\ and\
  \citenamefont {Dresselhaus}}]{Saito2016}%
  \BibitemOpen
  \bibfield  {author} {\bibinfo {author} {\bibfnamefont {R.}~\bibnamefont
  {Saito}}, \bibinfo {author} {\bibfnamefont {Y.}~\bibnamefont {Tatsumi}},
  \bibinfo {author} {\bibfnamefont {S.}~\bibnamefont {Huang}}, \bibinfo
  {author} {\bibfnamefont {X.}~\bibnamefont {Ling}},\ and\ \bibinfo {author}
  {\bibfnamefont {M.~S.}\ \bibnamefont {Dresselhaus}},\ }\bibfield  {title}
  {\bibinfo {title} {Raman spectroscopy of transition metal dichalcogenides},\
  }\href {https://doi.org/10.1088/0953-8984/28/35/353002} {\bibfield  {journal}
  {\bibinfo  {journal} {Journal of Physics: Condensed Matter}\ }\textbf
  {\bibinfo {volume} {28}},\ \bibinfo {pages} {353002} (\bibinfo {year}
  {2016})}\BibitemShut {NoStop}%
\bibitem [{\citenamefont {Lin}\ \emph {et~al.}(2017)\citenamefont {Lin},
  \citenamefont {Wu}, \citenamefont {Liu},\ and\ \citenamefont
  {Tan}}]{Lin2017}%
  \BibitemOpen
  \bibfield  {author} {\bibinfo {author} {\bibfnamefont {M.}~\bibnamefont
  {Lin}}, \bibinfo {author} {\bibfnamefont {J.}~\bibnamefont {Wu}}, \bibinfo
  {author} {\bibfnamefont {X.}~\bibnamefont {Liu}},\ and\ \bibinfo {author}
  {\bibfnamefont {P.}~\bibnamefont {Tan}},\ }\bibfield  {title} {\bibinfo
  {title} {Probing the shear and layer breathing modes in multilayer graphene
  by raman spectroscopy},\ }\href {https://doi.org/10.1002/jrs.5224} {\bibfield
   {journal} {\bibinfo  {journal} {Journal of Raman Spectroscopy}\ }\textbf
  {\bibinfo {volume} {49}},\ \bibinfo {pages} {19} (\bibinfo {year}
  {2017})}\BibitemShut {NoStop}%
\bibitem [{\citenamefont {Lui}\ \emph {et~al.}(2014)\citenamefont {Lui},
  \citenamefont {Ye}, \citenamefont {Keiser}, \citenamefont {Xiao},\ and\
  \citenamefont {He}}]{Lui2014}%
  \BibitemOpen
  \bibfield  {author} {\bibinfo {author} {\bibfnamefont {C.~H.}\ \bibnamefont
  {Lui}}, \bibinfo {author} {\bibfnamefont {Z.}~\bibnamefont {Ye}}, \bibinfo
  {author} {\bibfnamefont {C.}~\bibnamefont {Keiser}}, \bibinfo {author}
  {\bibfnamefont {X.}~\bibnamefont {Xiao}},\ and\ \bibinfo {author}
  {\bibfnamefont {R.}~\bibnamefont {He}},\ }\bibfield  {title} {\bibinfo
  {title} {Temperature-activated layer-breathing vibrations in few-layer
  graphene},\ }\href {https://doi.org/10.1021/nl501678j} {\bibfield  {journal}
  {\bibinfo  {journal} {Nano Letters}\ }\textbf {\bibinfo {volume} {14}},\
  \bibinfo {pages} {4615} (\bibinfo {year} {2014})}\BibitemShut {NoStop}%
\bibitem [{\citenamefont {Lui}\ and\ \citenamefont {Heinz}(2013)}]{Lui2013}%
  \BibitemOpen
  \bibfield  {author} {\bibinfo {author} {\bibfnamefont {C.~H.}\ \bibnamefont
  {Lui}}\ and\ \bibinfo {author} {\bibfnamefont {T.~F.}\ \bibnamefont
  {Heinz}},\ }\bibfield  {title} {\bibinfo {title} {Measurement of layer
  breathing mode vibrations in few-layer graphene},\ }\href
  {https://doi.org/10.1103/physrevb.87.121404} {\bibfield  {journal} {\bibinfo
  {journal} {Physical Review B}\ }\textbf {\bibinfo {volume} {87}},\ \bibinfo
  {pages} {121404} (\bibinfo {year} {2013})}\BibitemShut {NoStop}%
\bibitem [{\citenamefont {Lui}\ \emph {et~al.}(2015)\citenamefont {Lui},
  \citenamefont {Ye}, \citenamefont {Keiser}, \citenamefont {Barros},\ and\
  \citenamefont {He}}]{Lui2015}%
  \BibitemOpen
  \bibfield  {author} {\bibinfo {author} {\bibfnamefont {C.~H.}\ \bibnamefont
  {Lui}}, \bibinfo {author} {\bibfnamefont {Z.}~\bibnamefont {Ye}}, \bibinfo
  {author} {\bibfnamefont {C.}~\bibnamefont {Keiser}}, \bibinfo {author}
  {\bibfnamefont {E.~B.}\ \bibnamefont {Barros}},\ and\ \bibinfo {author}
  {\bibfnamefont {R.}~\bibnamefont {He}},\ }\bibfield  {title} {\bibinfo
  {title} {Stacking-dependent shear modes in trilayer graphene},\ }\bibfield
  {journal} {\bibinfo  {journal} {Applied Physics Letters}\ }\textbf {\bibinfo
  {volume} {106}},\ \href {https://doi.org/10.1063/1.4906579}
  {10.1063/1.4906579} (\bibinfo {year} {2015})\BibitemShut {NoStop}%
\bibitem [{\citenamefont {Zhao}\ \emph {et~al.}(2013)\citenamefont {Zhao},
  \citenamefont {Luo}, \citenamefont {Li}, \citenamefont {Zhang}, \citenamefont
  {Araujo}, \citenamefont {Gan}, \citenamefont {Wu}, \citenamefont {Zhang},
  \citenamefont {Quek}, \citenamefont {Dresselhaus},\ and\ \citenamefont
  {Xiong}}]{Zhao2013}%
  \BibitemOpen
  \bibfield  {author} {\bibinfo {author} {\bibfnamefont {Y.}~\bibnamefont
  {Zhao}}, \bibinfo {author} {\bibfnamefont {X.}~\bibnamefont {Luo}}, \bibinfo
  {author} {\bibfnamefont {H.}~\bibnamefont {Li}}, \bibinfo {author}
  {\bibfnamefont {J.}~\bibnamefont {Zhang}}, \bibinfo {author} {\bibfnamefont
  {P.~T.}\ \bibnamefont {Araujo}}, \bibinfo {author} {\bibfnamefont {C.~K.}\
  \bibnamefont {Gan}}, \bibinfo {author} {\bibfnamefont {J.}~\bibnamefont
  {Wu}}, \bibinfo {author} {\bibfnamefont {H.}~\bibnamefont {Zhang}}, \bibinfo
  {author} {\bibfnamefont {S.~Y.}\ \bibnamefont {Quek}}, \bibinfo {author}
  {\bibfnamefont {M.~S.}\ \bibnamefont {Dresselhaus}},\ and\ \bibinfo {author}
  {\bibfnamefont {Q.}~\bibnamefont {Xiong}},\ }\bibfield  {title} {\bibinfo
  {title} {Interlayer breathing and shear modes in few-trilayer {M}o{S}$_2$ and
  {WS}e$_2$},\ }\href {https://doi.org/10.1021/nl304169w} {\bibfield  {journal}
  {\bibinfo  {journal} {Nano Letters}\ }\textbf {\bibinfo {volume} {13}},\
  \bibinfo {pages} {1007} (\bibinfo {year} {2013})}\BibitemShut {NoStop}%
\bibitem [{\citenamefont {Chen}\ \emph {et~al.}(2015)\citenamefont {Chen},
  \citenamefont {Zheng}, \citenamefont {Fuhrer},\ and\ \citenamefont
  {Yan}}]{Chen2015}%
  \BibitemOpen
  \bibfield  {author} {\bibinfo {author} {\bibfnamefont {S.-Y.}\ \bibnamefont
  {Chen}}, \bibinfo {author} {\bibfnamefont {C.}~\bibnamefont {Zheng}},
  \bibinfo {author} {\bibfnamefont {M.~S.}\ \bibnamefont {Fuhrer}},\ and\
  \bibinfo {author} {\bibfnamefont {J.}~\bibnamefont {Yan}},\ }\bibfield
  {title} {\bibinfo {title} {Helicity-resolved raman scattering of mos2, mose2,
  ws2, and wse2 atomic layers},\ }\href
  {https://doi.org/10.1021/acs.nanolett.5b00092} {\bibfield  {journal}
  {\bibinfo  {journal} {Nano Letters}\ }\textbf {\bibinfo {volume} {15}},\
  \bibinfo {pages} {2526} (\bibinfo {year} {2015})}\BibitemShut {NoStop}%
\bibitem [{\citenamefont {van Baren}\ \emph {et~al.}(2019)\citenamefont {van
  Baren}, \citenamefont {Ye}, \citenamefont {Yan}, \citenamefont {Ye},
  \citenamefont {Rezaie}, \citenamefont {Yu}, \citenamefont {Liu},
  \citenamefont {He},\ and\ \citenamefont {Lui}}]{Baren2019}%
  \BibitemOpen
  \bibfield  {author} {\bibinfo {author} {\bibfnamefont {J.}~\bibnamefont {van
  Baren}}, \bibinfo {author} {\bibfnamefont {G.}~\bibnamefont {Ye}}, \bibinfo
  {author} {\bibfnamefont {J.-A.}\ \bibnamefont {Yan}}, \bibinfo {author}
  {\bibfnamefont {Z.}~\bibnamefont {Ye}}, \bibinfo {author} {\bibfnamefont
  {P.}~\bibnamefont {Rezaie}}, \bibinfo {author} {\bibfnamefont
  {P.}~\bibnamefont {Yu}}, \bibinfo {author} {\bibfnamefont {Z.}~\bibnamefont
  {Liu}}, \bibinfo {author} {\bibfnamefont {R.}~\bibnamefont {He}},\ and\
  \bibinfo {author} {\bibfnamefont {C.~H.}\ \bibnamefont {Lui}},\ }\bibfield
  {title} {\bibinfo {title} {Stacking-dependent interlayer phonons in 3r and 2h
  mos 2},\ }\href {https://doi.org/10.1088/2053-1583/ab0196} {\bibfield
  {journal} {\bibinfo  {journal} {2D Materials}\ }\textbf {\bibinfo {volume}
  {6}},\ \bibinfo {pages} {025022} (\bibinfo {year} {2019})}\BibitemShut
  {NoStop}%
\bibitem [{\citenamefont {Boora}\ \emph {et~al.}(2024)\citenamefont {Boora},
  \citenamefont {Lin}, \citenamefont {Chen}, \citenamefont {Trainor},
  \citenamefont {Robinson}, \citenamefont {Redwing},\ and\ \citenamefont
  {Suh}}]{Boora2024}%
  \BibitemOpen
  \bibfield  {author} {\bibinfo {author} {\bibfnamefont {M.}~\bibnamefont
  {Boora}}, \bibinfo {author} {\bibfnamefont {Y.-C.}\ \bibnamefont {Lin}},
  \bibinfo {author} {\bibfnamefont {C.}~\bibnamefont {Chen}}, \bibinfo {author}
  {\bibfnamefont {N.}~\bibnamefont {Trainor}}, \bibinfo {author} {\bibfnamefont
  {J.~A.}\ \bibnamefont {Robinson}}, \bibinfo {author} {\bibfnamefont {J.~M.}\
  \bibnamefont {Redwing}},\ and\ \bibinfo {author} {\bibfnamefont {J.~Y.}\
  \bibnamefont {Suh}},\ }\bibfield  {title} {\bibinfo {title} {Low-frequency
  raman study of large-area twisted bilayers of {WS}$_2$ stacked by an
  etchant-free transfer method},\ }\href
  {https://doi.org/10.1021/acsami.3c14708} {\bibfield  {journal} {\bibinfo
  {journal} {ACS Applied Materials \&; Interfaces}\ }\textbf {\bibinfo {volume}
  {16}},\ \bibinfo {pages} {2902} (\bibinfo {year} {2024})}\BibitemShut
  {NoStop}%
\bibitem [{\citenamefont {Cao}\ \emph {et~al.}(2018{\natexlab{a}})\citenamefont
  {Cao}, \citenamefont {Fatemi}, \citenamefont {Fang}, \citenamefont
  {Watanabe}, \citenamefont {Taniguchi}, \citenamefont {Kaxiras},\ and\
  \citenamefont {Jarillo-Herrero}}]{cao2018unconventional}%
  \BibitemOpen
  \bibfield  {author} {\bibinfo {author} {\bibfnamefont {Y.}~\bibnamefont
  {Cao}}, \bibinfo {author} {\bibfnamefont {V.}~\bibnamefont {Fatemi}},
  \bibinfo {author} {\bibfnamefont {S.}~\bibnamefont {Fang}}, \bibinfo {author}
  {\bibfnamefont {K.}~\bibnamefont {Watanabe}}, \bibinfo {author}
  {\bibfnamefont {T.}~\bibnamefont {Taniguchi}}, \bibinfo {author}
  {\bibfnamefont {E.}~\bibnamefont {Kaxiras}},\ and\ \bibinfo {author}
  {\bibfnamefont {P.}~\bibnamefont {Jarillo-Herrero}},\ }\bibfield  {title}
  {\bibinfo {title} {Unconventional superconductivity in magic-angle graphene
  superlattices},\ }\href {https://doi.org/10.1038/nature26160} {\bibfield
  {journal} {\bibinfo  {journal} {Nature}\ }\textbf {\bibinfo {volume} {556}},\
  \bibinfo {pages} {43} (\bibinfo {year} {2018}{\natexlab{a}})}\BibitemShut
  {NoStop}%
\bibitem [{\citenamefont {{Yankowitz}}\ \emph {et~al.}(2019)\citenamefont
  {{Yankowitz}}, \citenamefont {{Chen}}, \citenamefont {{Polshyn}},
  \citenamefont {{Zhang}}, \citenamefont {{Watanabe}}, \citenamefont
  {{Taniguchi}}, \citenamefont {{Graf}}, \citenamefont {{Young}},\ and\
  \citenamefont {{Dean}}}]{Yankowitz2019}%
  \BibitemOpen
  \bibfield  {author} {\bibinfo {author} {\bibfnamefont {M.}~\bibnamefont
  {{Yankowitz}}}, \bibinfo {author} {\bibfnamefont {S.}~\bibnamefont {{Chen}}},
  \bibinfo {author} {\bibfnamefont {H.}~\bibnamefont {{Polshyn}}}, \bibinfo
  {author} {\bibfnamefont {Y.}~\bibnamefont {{Zhang}}}, \bibinfo {author}
  {\bibfnamefont {K.}~\bibnamefont {{Watanabe}}}, \bibinfo {author}
  {\bibfnamefont {T.}~\bibnamefont {{Taniguchi}}}, \bibinfo {author}
  {\bibfnamefont {D.}~\bibnamefont {{Graf}}}, \bibinfo {author} {\bibfnamefont
  {A.~F.}\ \bibnamefont {{Young}}},\ and\ \bibinfo {author} {\bibfnamefont
  {C.~R.}\ \bibnamefont {{Dean}}},\ }\bibfield  {title} {\bibinfo {title}
  {{Tuning superconductivity in twisted bilayer graphene}},\ }\href
  {https://doi.org/10.1126/science.aav1910} {\bibfield  {journal} {\bibinfo
  {journal} {Science}\ }\textbf {\bibinfo {volume} {363}},\ \bibinfo {pages}
  {1059} (\bibinfo {year} {2019})}\BibitemShut {NoStop}%
\bibitem [{\citenamefont {Xia}\ \emph {et~al.}(2024)\citenamefont {Xia},
  \citenamefont {Han}, \citenamefont {Watanabe}, \citenamefont {Taniguchi},
  \citenamefont {Shan},\ and\ \citenamefont {Mak}}]{Xia2024}%
  \BibitemOpen
  \bibfield  {author} {\bibinfo {author} {\bibfnamefont {Y.}~\bibnamefont
  {Xia}}, \bibinfo {author} {\bibfnamefont {Z.}~\bibnamefont {Han}}, \bibinfo
  {author} {\bibfnamefont {K.}~\bibnamefont {Watanabe}}, \bibinfo {author}
  {\bibfnamefont {T.}~\bibnamefont {Taniguchi}}, \bibinfo {author}
  {\bibfnamefont {J.}~\bibnamefont {Shan}},\ and\ \bibinfo {author}
  {\bibfnamefont {K.~F.}\ \bibnamefont {Mak}},\ }\bibfield  {title} {\bibinfo
  {title} {Superconductivity in twisted bilayer wse2},\ }\href
  {https://doi.org/10.1038/s41586-024-08116-2} {\bibfield  {journal} {\bibinfo
  {journal} {Nature}\ }\textbf {\bibinfo {volume} {637}},\ \bibinfo {pages}
  {833} (\bibinfo {year} {2024})}\BibitemShut {NoStop}%
\bibitem [{\citenamefont {Guo}\ \emph {et~al.}(2025)\citenamefont {Guo},
  \citenamefont {Pack}, \citenamefont {Swann}, \citenamefont {Holtzman},
  \citenamefont {Cothrine}, \citenamefont {Watanabe}, \citenamefont
  {Taniguchi}, \citenamefont {Mandrus}, \citenamefont {Barmak}, \citenamefont
  {Hone}, \citenamefont {Millis}, \citenamefont {Pasupathy},\ and\
  \citenamefont {Dean}}]{Guo2025}%
  \BibitemOpen
  \bibfield  {author} {\bibinfo {author} {\bibfnamefont {Y.}~\bibnamefont
  {Guo}}, \bibinfo {author} {\bibfnamefont {J.}~\bibnamefont {Pack}}, \bibinfo
  {author} {\bibfnamefont {J.}~\bibnamefont {Swann}}, \bibinfo {author}
  {\bibfnamefont {L.}~\bibnamefont {Holtzman}}, \bibinfo {author}
  {\bibfnamefont {M.}~\bibnamefont {Cothrine}}, \bibinfo {author}
  {\bibfnamefont {K.}~\bibnamefont {Watanabe}}, \bibinfo {author}
  {\bibfnamefont {T.}~\bibnamefont {Taniguchi}}, \bibinfo {author}
  {\bibfnamefont {D.~G.}\ \bibnamefont {Mandrus}}, \bibinfo {author}
  {\bibfnamefont {K.}~\bibnamefont {Barmak}}, \bibinfo {author} {\bibfnamefont
  {J.}~\bibnamefont {Hone}}, \bibinfo {author} {\bibfnamefont {A.~J.}\
  \bibnamefont {Millis}}, \bibinfo {author} {\bibfnamefont {A.}~\bibnamefont
  {Pasupathy}},\ and\ \bibinfo {author} {\bibfnamefont {C.~R.}\ \bibnamefont
  {Dean}},\ }\bibfield  {title} {\bibinfo {title} {Superconductivity in 5.0°
  twisted bilayer wse2},\ }\href {https://doi.org/10.1038/s41586-024-08381-1}
  {\bibfield  {journal} {\bibinfo  {journal} {Nature}\ }\textbf {\bibinfo
  {volume} {637}},\ \bibinfo {pages} {839} (\bibinfo {year}
  {2025})}\BibitemShut {NoStop}%
\bibitem [{\citenamefont {Cao}\ \emph {et~al.}(2018{\natexlab{b}})\citenamefont
  {Cao}, \citenamefont {Fatemi}, \citenamefont {Demir}, \citenamefont {Fang},
  \citenamefont {Tomarken}, \citenamefont {Luo}, \citenamefont
  {Sanchez-Yamagishi}, \citenamefont {Watanabe}, \citenamefont {Taniguchi},
  \citenamefont {Kaxiras},\ and\ \citenamefont {et~al}}]{cao2018correlated}%
  \BibitemOpen
  \bibfield  {author} {\bibinfo {author} {\bibfnamefont {Y.}~\bibnamefont
  {Cao}}, \bibinfo {author} {\bibfnamefont {V.}~\bibnamefont {Fatemi}},
  \bibinfo {author} {\bibfnamefont {A.}~\bibnamefont {Demir}}, \bibinfo
  {author} {\bibfnamefont {S.}~\bibnamefont {Fang}}, \bibinfo {author}
  {\bibfnamefont {S.~L.}\ \bibnamefont {Tomarken}}, \bibinfo {author}
  {\bibfnamefont {J.~Y.}\ \bibnamefont {Luo}}, \bibinfo {author} {\bibfnamefont
  {J.~D.}\ \bibnamefont {Sanchez-Yamagishi}}, \bibinfo {author} {\bibfnamefont
  {K.}~\bibnamefont {Watanabe}}, \bibinfo {author} {\bibfnamefont
  {T.}~\bibnamefont {Taniguchi}}, \bibinfo {author} {\bibfnamefont
  {E.}~\bibnamefont {Kaxiras}},\ and\ \bibinfo {author} {\bibnamefont
  {et~al}},\ }\bibfield  {title} {\bibinfo {title} {Correlated insulator
  behaviour at half-filling in magic-angle graphene superlattices},\ }\href
  {https://doi.org/10.1038/nature26154} {\bibfield  {journal} {\bibinfo
  {journal} {Nature}\ }\textbf {\bibinfo {volume} {556}},\ \bibinfo {pages}
  {80} (\bibinfo {year} {2018}{\natexlab{b}})}\BibitemShut {NoStop}%
\bibitem [{\citenamefont {Kennes}\ \emph {et~al.}(2021)\citenamefont {Kennes},
  \citenamefont {Claassen}, \citenamefont {Xian}, \citenamefont {Georges},
  \citenamefont {Millis}, \citenamefont {Hone}, \citenamefont {Dean},
  \citenamefont {Basov}, \citenamefont {Pasupathy},\ and\ \citenamefont
  {Rubio}}]{Kennes2021}%
  \BibitemOpen
  \bibfield  {author} {\bibinfo {author} {\bibfnamefont {D.~M.}\ \bibnamefont
  {Kennes}}, \bibinfo {author} {\bibfnamefont {M.}~\bibnamefont {Claassen}},
  \bibinfo {author} {\bibfnamefont {L.}~\bibnamefont {Xian}}, \bibinfo {author}
  {\bibfnamefont {A.}~\bibnamefont {Georges}}, \bibinfo {author} {\bibfnamefont
  {A.~J.}\ \bibnamefont {Millis}}, \bibinfo {author} {\bibfnamefont
  {J.}~\bibnamefont {Hone}}, \bibinfo {author} {\bibfnamefont {C.~R.}\
  \bibnamefont {Dean}}, \bibinfo {author} {\bibfnamefont {D.~N.}\ \bibnamefont
  {Basov}}, \bibinfo {author} {\bibfnamefont {A.~N.}\ \bibnamefont
  {Pasupathy}},\ and\ \bibinfo {author} {\bibfnamefont {A.}~\bibnamefont
  {Rubio}},\ }\bibfield  {title} {\bibinfo {title} {Moiré heterostructures as
  a condensed-matter quantum simulator},\ }\href
  {https://doi.org/10.1038/s41567-020-01154-3} {\bibfield  {journal} {\bibinfo
  {journal} {Nature Physics}\ }\textbf {\bibinfo {volume} {17}},\ \bibinfo
  {pages} {155} (\bibinfo {year} {2021})}\BibitemShut {NoStop}%
\bibitem [{\citenamefont {Foutty}\ \emph {et~al.}(2023)\citenamefont {Foutty},
  \citenamefont {Yu}, \citenamefont {Devakul}, \citenamefont {Kometter},
  \citenamefont {Zhang}, \citenamefont {Watanabe}, \citenamefont {Taniguchi},
  \citenamefont {Fu},\ and\ \citenamefont {Feldman}}]{Foutty2023}%
  \BibitemOpen
  \bibfield  {author} {\bibinfo {author} {\bibfnamefont {B.~A.}\ \bibnamefont
  {Foutty}}, \bibinfo {author} {\bibfnamefont {J.}~\bibnamefont {Yu}}, \bibinfo
  {author} {\bibfnamefont {T.}~\bibnamefont {Devakul}}, \bibinfo {author}
  {\bibfnamefont {C.~R.}\ \bibnamefont {Kometter}}, \bibinfo {author}
  {\bibfnamefont {Y.}~\bibnamefont {Zhang}}, \bibinfo {author} {\bibfnamefont
  {K.}~\bibnamefont {Watanabe}}, \bibinfo {author} {\bibfnamefont
  {T.}~\bibnamefont {Taniguchi}}, \bibinfo {author} {\bibfnamefont
  {L.}~\bibnamefont {Fu}},\ and\ \bibinfo {author} {\bibfnamefont {B.~E.}\
  \bibnamefont {Feldman}},\ }\bibfield  {title} {\bibinfo {title} {Tunable spin
  and valley excitations of correlated insulators in $\gamma$-valley moiré
  bands},\ }\href {https://doi.org/10.1038/s41563-023-01534-z} {\bibfield
  {journal} {\bibinfo  {journal} {Nature Materials}\ }\textbf {\bibinfo
  {volume} {22}},\ \bibinfo {pages} {731} (\bibinfo {year} {2023})}\BibitemShut
  {NoStop}%
\bibitem [{\citenamefont {Campbell}\ \emph {et~al.}(2024)\citenamefont
  {Campbell}, \citenamefont {Vitale}, \citenamefont {Brotons-Gisbert},
  \citenamefont {Baek}, \citenamefont {Borel}, \citenamefont {Ivanova},
  \citenamefont {Taniguchi}, \citenamefont {Watanabe}, \citenamefont
  {Lischner},\ and\ \citenamefont {Gerardot}}]{Campbell2024}%
  \BibitemOpen
  \bibfield  {author} {\bibinfo {author} {\bibfnamefont {A.~J.}\ \bibnamefont
  {Campbell}}, \bibinfo {author} {\bibfnamefont {V.}~\bibnamefont {Vitale}},
  \bibinfo {author} {\bibfnamefont {M.}~\bibnamefont {Brotons-Gisbert}},
  \bibinfo {author} {\bibfnamefont {H.}~\bibnamefont {Baek}}, \bibinfo {author}
  {\bibfnamefont {A.}~\bibnamefont {Borel}}, \bibinfo {author} {\bibfnamefont
  {T.~V.}\ \bibnamefont {Ivanova}}, \bibinfo {author} {\bibfnamefont
  {T.}~\bibnamefont {Taniguchi}}, \bibinfo {author} {\bibfnamefont
  {K.}~\bibnamefont {Watanabe}}, \bibinfo {author} {\bibfnamefont
  {J.}~\bibnamefont {Lischner}},\ and\ \bibinfo {author} {\bibfnamefont
  {B.~D.}\ \bibnamefont {Gerardot}},\ }\bibfield  {title} {\bibinfo {title}
  {The interplay of field-tunable strongly correlated states in a multi-orbital
  moiré system},\ }\href {https://doi.org/10.1038/s41567-024-02385-4}
  {\bibfield  {journal} {\bibinfo  {journal} {Nature Physics}\ }\textbf
  {\bibinfo {volume} {20}},\ \bibinfo {pages} {589} (\bibinfo {year}
  {2024})}\BibitemShut {NoStop}%
\bibitem [{\citenamefont {Li}\ \emph {et~al.}(2021)\citenamefont {Li},
  \citenamefont {Hu}, \citenamefont {Feng}, \citenamefont {Zhou}, \citenamefont
  {An}, \citenamefont {Law}, \citenamefont {Wang},\ and\ \citenamefont
  {Lin}}]{Li2021a}%
  \BibitemOpen
  \bibfield  {author} {\bibinfo {author} {\bibfnamefont {E.}~\bibnamefont
  {Li}}, \bibinfo {author} {\bibfnamefont {J.-X.}\ \bibnamefont {Hu}}, \bibinfo
  {author} {\bibfnamefont {X.}~\bibnamefont {Feng}}, \bibinfo {author}
  {\bibfnamefont {Z.}~\bibnamefont {Zhou}}, \bibinfo {author} {\bibfnamefont
  {L.}~\bibnamefont {An}}, \bibinfo {author} {\bibfnamefont {K.~T.}\
  \bibnamefont {Law}}, \bibinfo {author} {\bibfnamefont {N.}~\bibnamefont
  {Wang}},\ and\ \bibinfo {author} {\bibfnamefont {N.}~\bibnamefont {Lin}},\
  }\bibfield  {title} {\bibinfo {title} {Lattice reconstruction induced
  multiple ultra-flat bands in twisted bilayer wse2},\ }\bibfield  {journal}
  {\bibinfo  {journal} {Nature Communications}\ }\textbf {\bibinfo {volume}
  {12}},\ \href {https://doi.org/10.1038/s41467-021-25924-6}
  {10.1038/s41467-021-25924-6} (\bibinfo {year} {2021})\BibitemShut {NoStop}%
\bibitem [{\citenamefont {Bistritzer}\ and\ \citenamefont
  {MacDonald}(2011)}]{Bistritzer2011}%
  \BibitemOpen
  \bibfield  {author} {\bibinfo {author} {\bibfnamefont {R.}~\bibnamefont
  {Bistritzer}}\ and\ \bibinfo {author} {\bibfnamefont {A.~H.}\ \bibnamefont
  {MacDonald}},\ }\bibfield  {title} {\bibinfo {title} {Moir{\'{e}} bands in
  twisted double-layer graphene},\ }\href
  {https://doi.org/10.1073/pnas.1108174108} {\bibfield  {journal} {\bibinfo
  {journal} {Proceedings of the National Academy of Sciences}\ }\textbf
  {\bibinfo {volume} {108}},\ \bibinfo {pages} {12233} (\bibinfo {year}
  {2011})}\BibitemShut {NoStop}%
\bibitem [{\citenamefont {Garcia-Ruiz}\ \emph {et~al.}(2021)\citenamefont
  {Garcia-Ruiz}, \citenamefont {Deng}, \citenamefont {Enaldiev},\ and\
  \citenamefont {Fal'ko}}]{AitorPRB2021}%
  \BibitemOpen
  \bibfield  {author} {\bibinfo {author} {\bibfnamefont {A.}~\bibnamefont
  {Garcia-Ruiz}}, \bibinfo {author} {\bibfnamefont {H.-Y.}\ \bibnamefont
  {Deng}}, \bibinfo {author} {\bibfnamefont {V.~V.}\ \bibnamefont {Enaldiev}},\
  and\ \bibinfo {author} {\bibfnamefont {V.~I.}\ \bibnamefont {Fal'ko}},\
  }\bibfield  {title} {\bibinfo {title} {Full slonczewski-weiss-mcclure
  parametrization of few-layer twistronic graphene},\ }\href
  {https://doi.org/10.1103/PhysRevB.104.085402} {\bibfield  {journal} {\bibinfo
   {journal} {Phys. Rev. B}\ }\textbf {\bibinfo {volume} {104}},\ \bibinfo
  {pages} {085402} (\bibinfo {year} {2021})}\BibitemShut {NoStop}%
\bibitem [{\citenamefont {Naik}\ \emph {et~al.}(2020)\citenamefont {Naik},
  \citenamefont {Kundu}, \citenamefont {Maity},\ and\ \citenamefont
  {Jain}}]{Naik2020}%
  \BibitemOpen
  \bibfield  {author} {\bibinfo {author} {\bibfnamefont {M.~H.}\ \bibnamefont
  {Naik}}, \bibinfo {author} {\bibfnamefont {S.}~\bibnamefont {Kundu}},
  \bibinfo {author} {\bibfnamefont {I.}~\bibnamefont {Maity}},\ and\ \bibinfo
  {author} {\bibfnamefont {M.}~\bibnamefont {Jain}},\ }\bibfield  {title}
  {\bibinfo {title} {Origin and evolution of ultraflat bands in twisted bilayer
  transition metal dichalcogenides: Realization of triangular quantum dots},\
  }\href {https://doi.org/10.1103/physrevb.102.075413} {\bibfield  {journal}
  {\bibinfo  {journal} {Physical Review B}\ }\textbf {\bibinfo {volume}
  {102}},\ \bibinfo {pages} {075413} (\bibinfo {year} {2020})}\BibitemShut
  {NoStop}%
\bibitem [{\citenamefont {Magorrian}\ \emph {et~al.}(2021)\citenamefont
  {Magorrian}, \citenamefont {Enaldiev}, \citenamefont {Z\'olyomi},
  \citenamefont {Ferreira}, \citenamefont {Fal’ko},\ and\ \citenamefont
  {Ruiz-Tijerina}}]{Magorrian2021}%
  \BibitemOpen
  \bibfield  {author} {\bibinfo {author} {\bibfnamefont {S.~J.}\ \bibnamefont
  {Magorrian}}, \bibinfo {author} {\bibfnamefont {V.~V.}\ \bibnamefont
  {Enaldiev}}, \bibinfo {author} {\bibfnamefont {V.}~\bibnamefont {Z\'olyomi}},
  \bibinfo {author} {\bibfnamefont {F.}~\bibnamefont {Ferreira}}, \bibinfo
  {author} {\bibfnamefont {V.~I.}\ \bibnamefont {Fal’ko}},\ and\ \bibinfo
  {author} {\bibfnamefont {D.~A.}\ \bibnamefont {Ruiz-Tijerina}},\ }\bibfield
  {title} {\bibinfo {title} {Multifaceted moiré superlattice physics in
  twisted {WS}e$_2$ bilayers},\ }\href
  {https://doi.org/10.1103/physrevb.104.125440} {\bibfield  {journal} {\bibinfo
   {journal} {Physical Review B}\ }\textbf {\bibinfo {volume} {104}},\ \bibinfo
  {pages} {125440} (\bibinfo {year} {2021})}\BibitemShut {NoStop}%
\bibitem [{\citenamefont {Zhang}\ \emph {et~al.}(2020)\citenamefont {Zhang},
  \citenamefont {Wang}, \citenamefont {Watanabe}, \citenamefont {Taniguchi},
  \citenamefont {Ueno}, \citenamefont {Tutuc},\ and\ \citenamefont
  {LeRoy}}]{Zhang2020}%
  \BibitemOpen
  \bibfield  {author} {\bibinfo {author} {\bibfnamefont {Z.}~\bibnamefont
  {Zhang}}, \bibinfo {author} {\bibfnamefont {Y.}~\bibnamefont {Wang}},
  \bibinfo {author} {\bibfnamefont {K.}~\bibnamefont {Watanabe}}, \bibinfo
  {author} {\bibfnamefont {T.}~\bibnamefont {Taniguchi}}, \bibinfo {author}
  {\bibfnamefont {K.}~\bibnamefont {Ueno}}, \bibinfo {author} {\bibfnamefont
  {E.}~\bibnamefont {Tutuc}},\ and\ \bibinfo {author} {\bibfnamefont {B.~J.}\
  \bibnamefont {LeRoy}},\ }\bibfield  {title} {\bibinfo {title} {Flat bands in
  twisted bilayer transition metal dichalcogenides},\ }\href
  {https://doi.org/10.1038/s41567-020-0958-x} {\bibfield  {journal} {\bibinfo
  {journal} {Nature Physics}\ }\textbf {\bibinfo {volume} {16}},\ \bibinfo
  {pages} {1093} (\bibinfo {year} {2020})}\BibitemShut {NoStop}%
\bibitem [{\citenamefont {Cocemasov}\ \emph {et~al.}(2013)\citenamefont
  {Cocemasov}, \citenamefont {Nika},\ and\ \citenamefont
  {Balandin}}]{BalandinPRB2013}%
  \BibitemOpen
  \bibfield  {author} {\bibinfo {author} {\bibfnamefont {A.~I.}\ \bibnamefont
  {Cocemasov}}, \bibinfo {author} {\bibfnamefont {D.~L.}\ \bibnamefont
  {Nika}},\ and\ \bibinfo {author} {\bibfnamefont {A.~A.}\ \bibnamefont
  {Balandin}},\ }\bibfield  {title} {\bibinfo {title} {Phonons in twisted
  bilayer graphene},\ }\href {https://doi.org/10.1103/PhysRevB.88.035428}
  {\bibfield  {journal} {\bibinfo  {journal} {Phys. Rev. B}\ }\textbf {\bibinfo
  {volume} {88}},\ \bibinfo {pages} {035428} (\bibinfo {year}
  {2013})}\BibitemShut {NoStop}%
\bibitem [{\citenamefont {Song}\ \emph {et~al.}(2019)\citenamefont {Song},
  \citenamefont {Liu},\ and\ \citenamefont {Zhang}}]{Song2019}%
  \BibitemOpen
  \bibfield  {author} {\bibinfo {author} {\bibfnamefont {H.-Q.}\ \bibnamefont
  {Song}}, \bibinfo {author} {\bibfnamefont {Z.}~\bibnamefont {Liu}},\ and\
  \bibinfo {author} {\bibfnamefont {D.-B.}\ \bibnamefont {Zhang}},\ }\bibfield
  {title} {\bibinfo {title} {Interlayer vibration of twisted bilayer graphene:
  A first-principles study},\ }\href
  {https://doi.org/10.1016/j.physleta.2019.05.025} {\bibfield  {journal}
  {\bibinfo  {journal} {Physics Letters A}\ }\textbf {\bibinfo {volume}
  {383}},\ \bibinfo {pages} {2628} (\bibinfo {year} {2019})}\BibitemShut
  {NoStop}%
\bibitem [{\citenamefont {Koshino}\ and\ \citenamefont
  {Son}(2019)}]{KoshinoPRB2019}%
  \BibitemOpen
  \bibfield  {author} {\bibinfo {author} {\bibfnamefont {M.}~\bibnamefont
  {Koshino}}\ and\ \bibinfo {author} {\bibfnamefont {Y.-W.}\ \bibnamefont
  {Son}},\ }\bibfield  {title} {\bibinfo {title} {Moir\'e phonons in twisted
  bilayer graphene},\ }\href {https://doi.org/10.1103/PhysRevB.100.075416}
  {\bibfield  {journal} {\bibinfo  {journal} {Phys. Rev. B}\ }\textbf {\bibinfo
  {volume} {100}},\ \bibinfo {pages} {075416} (\bibinfo {year}
  {2019})}\BibitemShut {NoStop}%
\bibitem [{\citenamefont {Quan}\ \emph {et~al.}(2021)\citenamefont {Quan},
  \citenamefont {Linhart}, \citenamefont {Lin}, \citenamefont {Lee},
  \citenamefont {Zhu}, \citenamefont {Wang}, \citenamefont {Hsu}, \citenamefont
  {Choi}, \citenamefont {Embley}, \citenamefont {Young}, \citenamefont
  {Taniguchi}, \citenamefont {Watanabe}, \citenamefont {Shih}, \citenamefont
  {Lai}, \citenamefont {MacDonald}, \citenamefont {Tan}, \citenamefont
  {Libisch},\ and\ \citenamefont {Li}}]{Quan2021}%
  \BibitemOpen
  \bibfield  {author} {\bibinfo {author} {\bibfnamefont {J.}~\bibnamefont
  {Quan}}, \bibinfo {author} {\bibfnamefont {L.}~\bibnamefont {Linhart}},
  \bibinfo {author} {\bibfnamefont {M.-L.}\ \bibnamefont {Lin}}, \bibinfo
  {author} {\bibfnamefont {D.}~\bibnamefont {Lee}}, \bibinfo {author}
  {\bibfnamefont {J.}~\bibnamefont {Zhu}}, \bibinfo {author} {\bibfnamefont
  {C.-Y.}\ \bibnamefont {Wang}}, \bibinfo {author} {\bibfnamefont {W.-T.}\
  \bibnamefont {Hsu}}, \bibinfo {author} {\bibfnamefont {J.}~\bibnamefont
  {Choi}}, \bibinfo {author} {\bibfnamefont {J.}~\bibnamefont {Embley}},
  \bibinfo {author} {\bibfnamefont {C.}~\bibnamefont {Young}}, \bibinfo
  {author} {\bibfnamefont {T.}~\bibnamefont {Taniguchi}}, \bibinfo {author}
  {\bibfnamefont {K.}~\bibnamefont {Watanabe}}, \bibinfo {author}
  {\bibfnamefont {C.-K.}\ \bibnamefont {Shih}}, \bibinfo {author}
  {\bibfnamefont {K.}~\bibnamefont {Lai}}, \bibinfo {author} {\bibfnamefont
  {A.~H.}\ \bibnamefont {MacDonald}}, \bibinfo {author} {\bibfnamefont {P.-H.}\
  \bibnamefont {Tan}}, \bibinfo {author} {\bibfnamefont {F.}~\bibnamefont
  {Libisch}},\ and\ \bibinfo {author} {\bibfnamefont {X.}~\bibnamefont {Li}},\
  }\bibfield  {title} {\bibinfo {title} {Phonon renormalization in
  reconstructed mos2 moiré superlattices},\ }\href
  {https://doi.org/10.1038/s41563-021-00960-1} {\bibfield  {journal} {\bibinfo
  {journal} {Nature Materials}\ }\textbf {\bibinfo {volume} {20}},\ \bibinfo
  {pages} {1100} (\bibinfo {year} {2021})}\BibitemShut {NoStop}%
\bibitem [{\citenamefont {Maity}\ \emph {et~al.}(2022)\citenamefont {Maity},
  \citenamefont {Mostofi},\ and\ \citenamefont {Lischner}}]{Maity2022}%
  \BibitemOpen
  \bibfield  {author} {\bibinfo {author} {\bibfnamefont {I.}~\bibnamefont
  {Maity}}, \bibinfo {author} {\bibfnamefont {A.~A.}\ \bibnamefont {Mostofi}},\
  and\ \bibinfo {author} {\bibfnamefont {J.}~\bibnamefont {Lischner}},\
  }\bibfield  {title} {\bibinfo {title} {Chiral valley phonons and flat phonon
  bands in moiré materials},\ }\href
  {https://doi.org/10.1103/physrevb.105.l041408} {\bibfield  {journal}
  {\bibinfo  {journal} {Physical Review B}\ }\textbf {\bibinfo {volume}
  {105}},\ \bibinfo {pages} {l041408} (\bibinfo {year} {2022})}\BibitemShut
  {NoStop}%
\bibitem [{\citenamefont {Weston}\ \emph {et~al.}(2020)\citenamefont {Weston},
  \citenamefont {Zou}, \citenamefont {Enaldiev}, \citenamefont {Summerfield},
  \citenamefont {Clark}, \citenamefont {Z{\'o}lyomi}, \citenamefont {Graham},
  \citenamefont {Yelgel}, \citenamefont {Magorrian}, \citenamefont {Zhou},
  \citenamefont {Zultak}, \citenamefont {Hopkinson}, \citenamefont {Barinov},
  \citenamefont {Bointon}, \citenamefont {Kretinin}, \citenamefont {Wilson},
  \citenamefont {Beton}, \citenamefont {Fal'ko}, \citenamefont {Haigh},\ and\
  \citenamefont {Gorbachev}}]{Weston2020}%
  \BibitemOpen
  \bibfield  {author} {\bibinfo {author} {\bibfnamefont {A.}~\bibnamefont
  {Weston}}, \bibinfo {author} {\bibfnamefont {Y.}~\bibnamefont {Zou}},
  \bibinfo {author} {\bibfnamefont {V.}~\bibnamefont {Enaldiev}}, \bibinfo
  {author} {\bibfnamefont {A.}~\bibnamefont {Summerfield}}, \bibinfo {author}
  {\bibfnamefont {N.}~\bibnamefont {Clark}}, \bibinfo {author} {\bibfnamefont
  {V.}~\bibnamefont {Z{\'o}lyomi}}, \bibinfo {author} {\bibfnamefont
  {A.}~\bibnamefont {Graham}}, \bibinfo {author} {\bibfnamefont
  {C.}~\bibnamefont {Yelgel}}, \bibinfo {author} {\bibfnamefont
  {S.}~\bibnamefont {Magorrian}}, \bibinfo {author} {\bibfnamefont
  {M.}~\bibnamefont {Zhou}}, \bibinfo {author} {\bibfnamefont {J.}~\bibnamefont
  {Zultak}}, \bibinfo {author} {\bibfnamefont {D.}~\bibnamefont {Hopkinson}},
  \bibinfo {author} {\bibfnamefont {A.}~\bibnamefont {Barinov}}, \bibinfo
  {author} {\bibfnamefont {T.~H.}\ \bibnamefont {Bointon}}, \bibinfo {author}
  {\bibfnamefont {A.}~\bibnamefont {Kretinin}}, \bibinfo {author}
  {\bibfnamefont {N.~R.}\ \bibnamefont {Wilson}}, \bibinfo {author}
  {\bibfnamefont {P.~H.}\ \bibnamefont {Beton}}, \bibinfo {author}
  {\bibfnamefont {V.~I.}\ \bibnamefont {Fal'ko}}, \bibinfo {author}
  {\bibfnamefont {S.~J.}\ \bibnamefont {Haigh}},\ and\ \bibinfo {author}
  {\bibfnamefont {R.}~\bibnamefont {Gorbachev}},\ }\bibfield  {title} {\bibinfo
  {title} {Atomic reconstruction in twisted bilayers of transition metal
  dichalcogenides},\ }\href {https://doi.org/10.1038/s41565-020-0682-9}
  {\bibfield  {journal} {\bibinfo  {journal} {Nature Nanotechnology}\ }\textbf
  {\bibinfo {volume} {15}},\ \bibinfo {pages} {592} (\bibinfo {year}
  {2020})}\BibitemShut {NoStop}%
\bibitem [{\citenamefont {Rosenberger}\ \emph {et~al.}(2020)\citenamefont
  {Rosenberger}, \citenamefont {Chuang}, \citenamefont {Phillips},
  \citenamefont {Oleshko}, \citenamefont {McCreary}, \citenamefont {Sivaram},
  \citenamefont {Hellberg},\ and\ \citenamefont {Jonker}}]{rosenberger2020}%
  \BibitemOpen
  \bibfield  {author} {\bibinfo {author} {\bibfnamefont {M.~R.}\ \bibnamefont
  {Rosenberger}}, \bibinfo {author} {\bibfnamefont {H.-J.}\ \bibnamefont
  {Chuang}}, \bibinfo {author} {\bibfnamefont {M.}~\bibnamefont {Phillips}},
  \bibinfo {author} {\bibfnamefont {V.~P.}\ \bibnamefont {Oleshko}}, \bibinfo
  {author} {\bibfnamefont {K.~M.}\ \bibnamefont {McCreary}}, \bibinfo {author}
  {\bibfnamefont {S.~V.}\ \bibnamefont {Sivaram}}, \bibinfo {author}
  {\bibfnamefont {C.~S.}\ \bibnamefont {Hellberg}},\ and\ \bibinfo {author}
  {\bibfnamefont {B.~T.}\ \bibnamefont {Jonker}},\ }\bibfield  {title}
  {\bibinfo {title} {Twist angle-dependent atomic reconstruction and moir{\'e}
  patterns in transition metal dichalcogenide heterostructures},\ }\href
  {https://doi.org/10.1021/acsnano.0c00088} {\bibfield  {journal} {\bibinfo
  {journal} {ACS Nano}\ }\textbf {\bibinfo {volume} {14}},\ \bibinfo {pages}
  {4550} (\bibinfo {year} {2020})}\BibitemShut {NoStop}%
\bibitem [{\citenamefont {Carr}\ \emph {et~al.}(2018)\citenamefont {Carr},
  \citenamefont {Massatt}, \citenamefont {Torrisi}, \citenamefont {Cazeaux},
  \citenamefont {Luskin},\ and\ \citenamefont {Kaxiras}}]{CarrPRB2018}%
  \BibitemOpen
  \bibfield  {author} {\bibinfo {author} {\bibfnamefont {S.}~\bibnamefont
  {Carr}}, \bibinfo {author} {\bibfnamefont {D.}~\bibnamefont {Massatt}},
  \bibinfo {author} {\bibfnamefont {S.~B.}\ \bibnamefont {Torrisi}}, \bibinfo
  {author} {\bibfnamefont {P.}~\bibnamefont {Cazeaux}}, \bibinfo {author}
  {\bibfnamefont {M.}~\bibnamefont {Luskin}},\ and\ \bibinfo {author}
  {\bibfnamefont {E.}~\bibnamefont {Kaxiras}},\ }\bibfield  {title} {\bibinfo
  {title} {Relaxation and domain formation in incommensurate two-dimensional
  heterostructures},\ }\href {https://doi.org/10.1103/PhysRevB.98.224102}
  {\bibfield  {journal} {\bibinfo  {journal} {Phys. Rev. B}\ }\textbf {\bibinfo
  {volume} {98}},\ \bibinfo {pages} {224102} (\bibinfo {year}
  {2018})}\BibitemShut {NoStop}%
\bibitem [{\citenamefont {Engelke}\ \emph {et~al.}(2023)\citenamefont
  {Engelke}, \citenamefont {Yoo}, \citenamefont {Carr}, \citenamefont {Xu},
  \citenamefont {Cazeaux}, \citenamefont {Allen}, \citenamefont {Valdivia},
  \citenamefont {Luskin}, \citenamefont {Kaxiras}, \citenamefont {Kim},
  \citenamefont {Han},\ and\ \citenamefont {Kim}}]{Engelke2023}%
  \BibitemOpen
  \bibfield  {author} {\bibinfo {author} {\bibfnamefont {R.}~\bibnamefont
  {Engelke}}, \bibinfo {author} {\bibfnamefont {H.}~\bibnamefont {Yoo}},
  \bibinfo {author} {\bibfnamefont {S.}~\bibnamefont {Carr}}, \bibinfo {author}
  {\bibfnamefont {K.}~\bibnamefont {Xu}}, \bibinfo {author} {\bibfnamefont
  {P.}~\bibnamefont {Cazeaux}}, \bibinfo {author} {\bibfnamefont
  {R.}~\bibnamefont {Allen}}, \bibinfo {author} {\bibfnamefont {A.~M.}\
  \bibnamefont {Valdivia}}, \bibinfo {author} {\bibfnamefont {M.}~\bibnamefont
  {Luskin}}, \bibinfo {author} {\bibfnamefont {E.}~\bibnamefont {Kaxiras}},
  \bibinfo {author} {\bibfnamefont {M.}~\bibnamefont {Kim}}, \bibinfo {author}
  {\bibfnamefont {J.~H.}\ \bibnamefont {Han}},\ and\ \bibinfo {author}
  {\bibfnamefont {P.}~\bibnamefont {Kim}},\ }\bibfield  {title} {\bibinfo
  {title} {Topological nature of dislocation networks in two-dimensional
  moir{\'{e}} materials},\ }\href {https://doi.org/10.1103/physrevb.107.125413}
  {\bibfield  {journal} {\bibinfo  {journal} {Physical Review B}\ }\textbf
  {\bibinfo {volume} {107}},\ \bibinfo {pages} {125413} (\bibinfo {year}
  {2023})}\BibitemShut {NoStop}%
\bibitem [{\citenamefont {Weston}\ \emph {et~al.}(2022)\citenamefont {Weston},
  \citenamefont {Castanon}, \citenamefont {Enaldiev}, \citenamefont {Ferreira},
  \citenamefont {Bhattacharjee}, \citenamefont {Xu}, \citenamefont
  {Corte-Le{\'{o}}n}, \citenamefont {Wu}, \citenamefont {Clark}, \citenamefont
  {Summerfield}, \citenamefont {Hashimoto}, \citenamefont {Gao}, \citenamefont
  {Wang}, \citenamefont {Hamer}, \citenamefont {Read}, \citenamefont
  {Fumagalli}, \citenamefont {Kretinin}, \citenamefont {Haigh}, \citenamefont
  {Kazakova}, \citenamefont {Geim}, \citenamefont {Fal'ko},\ and\ \citenamefont
  {Gorbachev}}]{Weston2022}%
  \BibitemOpen
  \bibfield  {author} {\bibinfo {author} {\bibfnamefont {A.}~\bibnamefont
  {Weston}}, \bibinfo {author} {\bibfnamefont {E.~G.}\ \bibnamefont
  {Castanon}}, \bibinfo {author} {\bibfnamefont {V.}~\bibnamefont {Enaldiev}},
  \bibinfo {author} {\bibfnamefont {F.}~\bibnamefont {Ferreira}}, \bibinfo
  {author} {\bibfnamefont {S.}~\bibnamefont {Bhattacharjee}}, \bibinfo {author}
  {\bibfnamefont {S.}~\bibnamefont {Xu}}, \bibinfo {author} {\bibfnamefont
  {H.}~\bibnamefont {Corte-Le{\'{o}}n}}, \bibinfo {author} {\bibfnamefont
  {Z.}~\bibnamefont {Wu}}, \bibinfo {author} {\bibfnamefont {N.}~\bibnamefont
  {Clark}}, \bibinfo {author} {\bibfnamefont {A.}~\bibnamefont {Summerfield}},
  \bibinfo {author} {\bibfnamefont {T.}~\bibnamefont {Hashimoto}}, \bibinfo
  {author} {\bibfnamefont {Y.}~\bibnamefont {Gao}}, \bibinfo {author}
  {\bibfnamefont {W.}~\bibnamefont {Wang}}, \bibinfo {author} {\bibfnamefont
  {M.}~\bibnamefont {Hamer}}, \bibinfo {author} {\bibfnamefont
  {H.}~\bibnamefont {Read}}, \bibinfo {author} {\bibfnamefont {L.}~\bibnamefont
  {Fumagalli}}, \bibinfo {author} {\bibfnamefont {A.~V.}\ \bibnamefont
  {Kretinin}}, \bibinfo {author} {\bibfnamefont {S.~J.}\ \bibnamefont {Haigh}},
  \bibinfo {author} {\bibfnamefont {O.}~\bibnamefont {Kazakova}}, \bibinfo
  {author} {\bibfnamefont {A.~K.}\ \bibnamefont {Geim}}, \bibinfo {author}
  {\bibfnamefont {V.~I.}\ \bibnamefont {Fal'ko}},\ and\ \bibinfo {author}
  {\bibfnamefont {R.}~\bibnamefont {Gorbachev}},\ }\bibfield  {title} {\bibinfo
  {title} {Interfacial ferroelectricity in marginally twisted 2d
  semiconductors},\ }\bibfield  {journal} {\bibinfo  {journal} {Nature
  Nanotechnology}\ }\href {https://doi.org/10.1038/s41565-022-01072-w}
  {10.1038/s41565-022-01072-w} (\bibinfo {year} {2022})\BibitemShut {NoStop}%
\bibitem [{\citenamefont {Ferreira}\ \emph {et~al.}(2021)\citenamefont
  {Ferreira}, \citenamefont {Enaldiev}, \citenamefont {Fal'ko},\ and\
  \citenamefont {Magorrian}}]{Ferreira2021}%
  \BibitemOpen
  \bibfield  {author} {\bibinfo {author} {\bibfnamefont {F.}~\bibnamefont
  {Ferreira}}, \bibinfo {author} {\bibfnamefont {V.~V.}\ \bibnamefont
  {Enaldiev}}, \bibinfo {author} {\bibfnamefont {V.~I.}\ \bibnamefont
  {Fal'ko}},\ and\ \bibinfo {author} {\bibfnamefont {S.~J.}\ \bibnamefont
  {Magorrian}},\ }\bibfield  {title} {\bibinfo {title} {Weak ferroelectric
  charge transfer in layer-asymmetric bilayers of 2{D} semiconductors},\ }\href
  {https://doi.org/10.1038/s41598-021-92710-1} {\bibfield  {journal} {\bibinfo
  {journal} {Scientific Reports}\ }\textbf {\bibinfo {volume} {11}},\ \bibinfo
  {pages} {13422} (\bibinfo {year} {2021})}\BibitemShut {NoStop}%
\bibitem [{\citenamefont {Enaldiev}\ \emph {et~al.}(2022)\citenamefont
  {Enaldiev}, \citenamefont {Ferreira},\ and\ \citenamefont
  {Fal'ko}}]{Enaldiev2022}%
  \BibitemOpen
  \bibfield  {author} {\bibinfo {author} {\bibfnamefont {V.~V.}\ \bibnamefont
  {Enaldiev}}, \bibinfo {author} {\bibfnamefont {F.}~\bibnamefont {Ferreira}},\
  and\ \bibinfo {author} {\bibfnamefont {V.~I.}\ \bibnamefont {Fal'ko}},\
  }\bibfield  {title} {\bibinfo {title} {A scalable network model for
  electrically tunable ferroelectric domain structure in twistronic bilayers of
  two-dimensional semiconductors},\ }\href
  {https://doi.org/10.1021/acs.nanolett.1c04210} {\bibfield  {journal}
  {\bibinfo  {journal} {Nano Letters}\ }\textbf {\bibinfo {volume} {22}},\
  \bibinfo {pages} {1534} (\bibinfo {year} {2022})}\BibitemShut {NoStop}%
\bibitem [{\citenamefont {Ko}\ \emph {et~al.}(2023)\citenamefont {Ko},
  \citenamefont {Yuk}, \citenamefont {Engelke}, \citenamefont {Carr},
  \citenamefont {Kim}, \citenamefont {Park}, \citenamefont {Heo}, \citenamefont
  {Kim}, \citenamefont {Kim}, \citenamefont {Kim}, \citenamefont {Taniguchi},
  \citenamefont {Watanabe}, \citenamefont {Park}, \citenamefont {Kaxiras},
  \citenamefont {Yang}, \citenamefont {Kim},\ and\ \citenamefont
  {Yoo}}]{Ko2023}%
  \BibitemOpen
  \bibfield  {author} {\bibinfo {author} {\bibfnamefont {K.}~\bibnamefont
  {Ko}}, \bibinfo {author} {\bibfnamefont {A.}~\bibnamefont {Yuk}}, \bibinfo
  {author} {\bibfnamefont {R.}~\bibnamefont {Engelke}}, \bibinfo {author}
  {\bibfnamefont {S.}~\bibnamefont {Carr}}, \bibinfo {author} {\bibfnamefont
  {J.}~\bibnamefont {Kim}}, \bibinfo {author} {\bibfnamefont {D.}~\bibnamefont
  {Park}}, \bibinfo {author} {\bibfnamefont {H.}~\bibnamefont {Heo}}, \bibinfo
  {author} {\bibfnamefont {H.-M.}\ \bibnamefont {Kim}}, \bibinfo {author}
  {\bibfnamefont {S.-G.}\ \bibnamefont {Kim}}, \bibinfo {author} {\bibfnamefont
  {H.}~\bibnamefont {Kim}}, \bibinfo {author} {\bibfnamefont {T.}~\bibnamefont
  {Taniguchi}}, \bibinfo {author} {\bibfnamefont {K.}~\bibnamefont {Watanabe}},
  \bibinfo {author} {\bibfnamefont {H.}~\bibnamefont {Park}}, \bibinfo {author}
  {\bibfnamefont {E.}~\bibnamefont {Kaxiras}}, \bibinfo {author} {\bibfnamefont
  {S.~M.}\ \bibnamefont {Yang}}, \bibinfo {author} {\bibfnamefont
  {P.}~\bibnamefont {Kim}},\ and\ \bibinfo {author} {\bibfnamefont
  {H.}~\bibnamefont {Yoo}},\ }\bibfield  {title} {\bibinfo {title} {Operando
  electron microscopy investigation of polar domain dynamics in twisted van der
  waals homobilayers},\ }\href {https://doi.org/10.1038/s41563-023-01595-0}
  {\bibfield  {journal} {\bibinfo  {journal} {Nature Materials}\ }\textbf
  {\bibinfo {volume} {22}},\ \bibinfo {pages} {992} (\bibinfo {year}
  {2023})}\BibitemShut {NoStop}%
\bibitem [{\citenamefont {Molino}\ \emph {et~al.}(2023)\citenamefont {Molino},
  \citenamefont {Aggarwal}, \citenamefont {Enaldiev}, \citenamefont
  {Plumadore}, \citenamefont {I.~Fal´ko},\ and\ \citenamefont
  {Luican‐Mayer}}]{Molino2023}%
  \BibitemOpen
  \bibfield  {author} {\bibinfo {author} {\bibfnamefont {L.}~\bibnamefont
  {Molino}}, \bibinfo {author} {\bibfnamefont {L.}~\bibnamefont {Aggarwal}},
  \bibinfo {author} {\bibfnamefont {V.}~\bibnamefont {Enaldiev}}, \bibinfo
  {author} {\bibfnamefont {R.}~\bibnamefont {Plumadore}}, \bibinfo {author}
  {\bibfnamefont {V.}~\bibnamefont {I.~Fal´ko}},\ and\ \bibinfo {author}
  {\bibfnamefont {A.}~\bibnamefont {Luican‐Mayer}},\ }\bibfield  {title}
  {\bibinfo {title} {Ferroelectric switching at symmetry‐broken interfaces by
  local control of dislocations networks},\ }\bibfield  {journal} {\bibinfo
  {journal} {Advanced Materials}\ }\textbf {\bibinfo {volume} {35}},\ \href
  {https://doi.org/10.1002/adma.202207816} {10.1002/adma.202207816} (\bibinfo
  {year} {2023})\BibitemShut {NoStop}%
\bibitem [{\citenamefont {Ochoa}(2019)}]{PRB2019Ochoa}%
  \BibitemOpen
  \bibfield  {author} {\bibinfo {author} {\bibfnamefont {H.}~\bibnamefont
  {Ochoa}},\ }\bibfield  {title} {\bibinfo {title} {Moir\'e-pattern
  fluctuations and electron-phason coupling in twisted bilayer graphene},\
  }\href {https://doi.org/10.1103/PhysRevB.100.155426} {\bibfield  {journal}
  {\bibinfo  {journal} {Phys. Rev. B}\ }\textbf {\bibinfo {volume} {100}},\
  \bibinfo {pages} {155426} (\bibinfo {year} {2019})}\BibitemShut {NoStop}%
\bibitem [{\citenamefont {Maity}\ \emph {et~al.}(2020)\citenamefont {Maity},
  \citenamefont {Naik}, \citenamefont {Maiti}, \citenamefont {Krishnamurthy},\
  and\ \citenamefont {Jain}}]{Maity2020}%
  \BibitemOpen
  \bibfield  {author} {\bibinfo {author} {\bibfnamefont {I.}~\bibnamefont
  {Maity}}, \bibinfo {author} {\bibfnamefont {M.~H.}\ \bibnamefont {Naik}},
  \bibinfo {author} {\bibfnamefont {P.~K.}\ \bibnamefont {Maiti}}, \bibinfo
  {author} {\bibfnamefont {H.~R.}\ \bibnamefont {Krishnamurthy}},\ and\
  \bibinfo {author} {\bibfnamefont {M.}~\bibnamefont {Jain}},\ }\bibfield
  {title} {\bibinfo {title} {Phonons in twisted transition-metal dichalcogenide
  bilayers: Ultrasoft phasons and a transition from a superlubric to a pinned
  phase},\ }\href {https://doi.org/10.1103/physrevresearch.2.013335} {\bibfield
   {journal} {\bibinfo  {journal} {Physical Review Research}\ }\textbf
  {\bibinfo {volume} {2}},\ \bibinfo {pages} {013335} (\bibinfo {year}
  {2020})}\BibitemShut {NoStop}%
\bibitem [{\citenamefont {Samajdar}\ \emph {et~al.}(2022)\citenamefont
  {Samajdar}, \citenamefont {Teng},\ and\ \citenamefont
  {Scheurer}}]{PRB2022_Samajdar}%
  \BibitemOpen
  \bibfield  {author} {\bibinfo {author} {\bibfnamefont {R.}~\bibnamefont
  {Samajdar}}, \bibinfo {author} {\bibfnamefont {Y.}~\bibnamefont {Teng}},\
  and\ \bibinfo {author} {\bibfnamefont {M.~S.}\ \bibnamefont {Scheurer}},\
  }\bibfield  {title} {\bibinfo {title} {Moir\'e phonons and impact of
  electronic symmetry breaking in twisted trilayer graphene},\ }\href
  {https://doi.org/10.1103/PhysRevB.106.L201403} {\bibfield  {journal}
  {\bibinfo  {journal} {Phys. Rev. B}\ }\textbf {\bibinfo {volume} {106}},\
  \bibinfo {pages} {L201403} (\bibinfo {year} {2022})}\BibitemShut {NoStop}%
\bibitem [{\citenamefont {Liu}\ \emph {et~al.}(2022)\citenamefont {Liu},
  \citenamefont {Peng}, \citenamefont {Sun},\ and\ \citenamefont
  {Liu}}]{Liu2022}%
  \BibitemOpen
  \bibfield  {author} {\bibinfo {author} {\bibfnamefont {X.}~\bibnamefont
  {Liu}}, \bibinfo {author} {\bibfnamefont {R.}~\bibnamefont {Peng}}, \bibinfo
  {author} {\bibfnamefont {Z.}~\bibnamefont {Sun}},\ and\ \bibinfo {author}
  {\bibfnamefont {J.}~\bibnamefont {Liu}},\ }\bibfield  {title} {\bibinfo
  {title} {Moiré phonons in magic-angle twisted bilayer graphene},\ }\href
  {https://doi.org/10.1021/acs.nanolett.2c02010} {\bibfield  {journal}
  {\bibinfo  {journal} {Nano Letters}\ }\textbf {\bibinfo {volume} {22}},\
  \bibinfo {pages} {7791} (\bibinfo {year} {2022})}\BibitemShut {NoStop}%
\bibitem [{\citenamefont {de~Jong}\ \emph {et~al.}(2022)\citenamefont
  {de~Jong}, \citenamefont {Benschop}, \citenamefont {Chen}, \citenamefont
  {Krasovskii}, \citenamefont {de~Dood}, \citenamefont {Tromp}, \citenamefont
  {Allan},\ and\ \citenamefont {van~der Molen}}]{Jong2022}%
  \BibitemOpen
  \bibfield  {author} {\bibinfo {author} {\bibfnamefont {T.~A.}\ \bibnamefont
  {de~Jong}}, \bibinfo {author} {\bibfnamefont {T.}~\bibnamefont {Benschop}},
  \bibinfo {author} {\bibfnamefont {X.}~\bibnamefont {Chen}}, \bibinfo {author}
  {\bibfnamefont {E.~E.}\ \bibnamefont {Krasovskii}}, \bibinfo {author}
  {\bibfnamefont {M.~J.~A.}\ \bibnamefont {de~Dood}}, \bibinfo {author}
  {\bibfnamefont {R.~M.}\ \bibnamefont {Tromp}}, \bibinfo {author}
  {\bibfnamefont {M.~P.}\ \bibnamefont {Allan}},\ and\ \bibinfo {author}
  {\bibfnamefont {S.~J.}\ \bibnamefont {van~der Molen}},\ }\bibfield  {title}
  {\bibinfo {title} {Imaging moiré deformation and dynamics in twisted bilayer
  graphene},\ }\bibfield  {journal} {\bibinfo  {journal} {Nature
  Communications}\ }\textbf {\bibinfo {volume} {13}},\ \href
  {https://doi.org/10.1038/s41467-021-27646-1} {10.1038/s41467-021-27646-1}
  (\bibinfo {year} {2022})\BibitemShut {NoStop}%
\bibitem [{\citenamefont {Lu}\ \emph {et~al.}(2022)\citenamefont {Lu},
  \citenamefont {Zhu}, \citenamefont {Angeli}, \citenamefont {Larson},\ and\
  \citenamefont {Kaxiras}}]{Lu2022}%
  \BibitemOpen
  \bibfield  {author} {\bibinfo {author} {\bibfnamefont {J.~Z.}\ \bibnamefont
  {Lu}}, \bibinfo {author} {\bibfnamefont {Z.}~\bibnamefont {Zhu}}, \bibinfo
  {author} {\bibfnamefont {M.}~\bibnamefont {Angeli}}, \bibinfo {author}
  {\bibfnamefont {D.~T.}\ \bibnamefont {Larson}},\ and\ \bibinfo {author}
  {\bibfnamefont {E.}~\bibnamefont {Kaxiras}},\ }\bibfield  {title} {\bibinfo
  {title} {Low-energy moiré phonons in twisted bilayer van der waals
  heterostructures},\ }\href {https://doi.org/10.1103/physrevb.106.144305}
  {\bibfield  {journal} {\bibinfo  {journal} {Physical Review B}\ }\textbf
  {\bibinfo {volume} {106}},\ \bibinfo {pages} {144305} (\bibinfo {year}
  {2022})}\BibitemShut {NoStop}%
\bibitem [{\citenamefont {Cappelluti}\ \emph {et~al.}(2023)\citenamefont
  {Cappelluti}, \citenamefont {Silva-Guill\'en}, \citenamefont {Rostami},\ and\
  \citenamefont {Guinea}}]{PRB2023Cappelluti}%
  \BibitemOpen
  \bibfield  {author} {\bibinfo {author} {\bibfnamefont {E.}~\bibnamefont
  {Cappelluti}}, \bibinfo {author} {\bibfnamefont {J.~A.}\ \bibnamefont
  {Silva-Guill\'en}}, \bibinfo {author} {\bibfnamefont {H.}~\bibnamefont
  {Rostami}},\ and\ \bibinfo {author} {\bibfnamefont {F.}~\bibnamefont
  {Guinea}},\ }\bibfield  {title} {\bibinfo {title} {Flat-band optical phonons
  in twisted bilayer graphene},\ }\href
  {https://doi.org/10.1103/PhysRevB.108.125401} {\bibfield  {journal} {\bibinfo
   {journal} {Phys. Rev. B}\ }\textbf {\bibinfo {volume} {108}},\ \bibinfo
  {pages} {125401} (\bibinfo {year} {2023})}\BibitemShut {NoStop}%
\bibitem [{\citenamefont {Girotto}\ \emph {et~al.}(2023)\citenamefont
  {Girotto}, \citenamefont {Linhart},\ and\ \citenamefont
  {Libisch}}]{PRB2023Girotto}%
  \BibitemOpen
  \bibfield  {author} {\bibinfo {author} {\bibfnamefont {N.}~\bibnamefont
  {Girotto}}, \bibinfo {author} {\bibfnamefont {L.}~\bibnamefont {Linhart}},\
  and\ \bibinfo {author} {\bibfnamefont {F.}~\bibnamefont {Libisch}},\
  }\bibfield  {title} {\bibinfo {title} {Coupled phonons in twisted bilayer
  graphene},\ }\href {https://doi.org/10.1103/PhysRevB.108.155415} {\bibfield
  {journal} {\bibinfo  {journal} {Phys. Rev. B}\ }\textbf {\bibinfo {volume}
  {108}},\ \bibinfo {pages} {155415} (\bibinfo {year} {2023})}\BibitemShut
  {NoStop}%
\bibitem [{\citenamefont {Butz}\ \emph {et~al.}(2013)\citenamefont {Butz},
  \citenamefont {Dolle}, \citenamefont {Niekiel}, \citenamefont {Weber},
  \citenamefont {Waldmann}, \citenamefont {Weber}, \citenamefont {Meyer},\ and\
  \citenamefont {Spiecker}}]{Butz2013}%
  \BibitemOpen
  \bibfield  {author} {\bibinfo {author} {\bibfnamefont {B.}~\bibnamefont
  {Butz}}, \bibinfo {author} {\bibfnamefont {C.}~\bibnamefont {Dolle}},
  \bibinfo {author} {\bibfnamefont {F.}~\bibnamefont {Niekiel}}, \bibinfo
  {author} {\bibfnamefont {K.}~\bibnamefont {Weber}}, \bibinfo {author}
  {\bibfnamefont {D.}~\bibnamefont {Waldmann}}, \bibinfo {author}
  {\bibfnamefont {H.~B.}\ \bibnamefont {Weber}}, \bibinfo {author}
  {\bibfnamefont {B.}~\bibnamefont {Meyer}},\ and\ \bibinfo {author}
  {\bibfnamefont {E.}~\bibnamefont {Spiecker}},\ }\bibfield  {title} {\bibinfo
  {title} {Dislocations in bilayer graphene},\ }\href
  {https://doi.org/10.1038/nature12780} {\bibfield  {journal} {\bibinfo
  {journal} {Nature}\ }\textbf {\bibinfo {volume} {505}},\ \bibinfo {pages}
  {533} (\bibinfo {year} {2013})}\BibitemShut {NoStop}%
\bibitem [{\citenamefont {Liang}\ \emph {et~al.}(2023)\citenamefont {Liang},
  \citenamefont {Yang}, \citenamefont {Xiao}, \citenamefont {Chen},
  \citenamefont {Dadap}, \citenamefont {Rottler},\ and\ \citenamefont
  {Ye}}]{Liang2023}%
  \BibitemOpen
  \bibfield  {author} {\bibinfo {author} {\bibfnamefont {J.}~\bibnamefont
  {Liang}}, \bibinfo {author} {\bibfnamefont {D.}~\bibnamefont {Yang}},
  \bibinfo {author} {\bibfnamefont {Y.}~\bibnamefont {Xiao}}, \bibinfo {author}
  {\bibfnamefont {S.}~\bibnamefont {Chen}}, \bibinfo {author} {\bibfnamefont
  {J.~I.}\ \bibnamefont {Dadap}}, \bibinfo {author} {\bibfnamefont
  {J.}~\bibnamefont {Rottler}},\ and\ \bibinfo {author} {\bibfnamefont
  {Z.}~\bibnamefont {Ye}},\ }\bibfield  {title} {\bibinfo {title} {Shear
  strain-induced two-dimensional slip avalanches in rhombohedral {MoS}$_2$},\
  }\href {https://doi.org/10.1021/acs.nanolett.3c01487} {\bibfield  {journal}
  {\bibinfo  {journal} {Nano Letters}\ }\textbf {\bibinfo {volume} {23}},\
  \bibinfo {pages} {7228} (\bibinfo {year} {2023})}\BibitemShut {NoStop}%
\bibitem [{\citenamefont {Enaldiev}(2025)}]{Enaldiev2025}%
  \BibitemOpen
  \bibfield  {author} {\bibinfo {author} {\bibfnamefont {V.~V.}\ \bibnamefont
  {Enaldiev}},\ }\bibfield  {title} {\bibinfo {title} {Long-wavelength
  interdomain phonons and instability of dislocations in small-angle twisted
  bilayers},\ }\href {https://doi.org/10.1088/2053-1583/adac6f} {\bibfield
  {journal} {\bibinfo  {journal} {2D Materials}\ }\textbf {\bibinfo {volume}
  {12}},\ \bibinfo {pages} {021001} (\bibinfo {year} {2025})}\BibitemShut
  {NoStop}%
\bibitem [{\citenamefont {Zhou}\ \emph {et~al.}(2015)\citenamefont {Zhou},
  \citenamefont {Han}, \citenamefont {Dai}, \citenamefont {Sun},\ and\
  \citenamefont {Srolovitz}}]{Zhou2015}%
  \BibitemOpen
  \bibfield  {author} {\bibinfo {author} {\bibfnamefont {S.}~\bibnamefont
  {Zhou}}, \bibinfo {author} {\bibfnamefont {J.}~\bibnamefont {Han}}, \bibinfo
  {author} {\bibfnamefont {S.}~\bibnamefont {Dai}}, \bibinfo {author}
  {\bibfnamefont {J.}~\bibnamefont {Sun}},\ and\ \bibinfo {author}
  {\bibfnamefont {D.~J.}\ \bibnamefont {Srolovitz}},\ }\bibfield  {title}
  {\bibinfo {title} {van der waals bilayer energetics: Generalized
  stacking-fault energy of graphene, boron nitride, and graphene/boron nitride
  bilayers},\ }\href {https://doi.org/10.1103/physrevb.92.155438} {\bibfield
  {journal} {\bibinfo  {journal} {Physical Review B}\ }\textbf {\bibinfo
  {volume} {92}},\ \bibinfo {pages} {155438} (\bibinfo {year}
  {2015})}\BibitemShut {NoStop}%
\bibitem [{\citenamefont {Enaldiev}(2024)}]{Enaldiev2024}%
  \BibitemOpen
  \bibfield  {author} {\bibinfo {author} {\bibfnamefont {V.}~\bibnamefont
  {Enaldiev}},\ }\bibfield  {title} {\bibinfo {title} {Dislocations in
  twistronic heterostructures},\ }\href
  {https://doi.org/10.1088/2053-1583/ad3b13} {\bibfield  {journal} {\bibinfo
  {journal} {2D Materials}\ }\textbf {\bibinfo {volume} {11}},\ \bibinfo
  {pages} {035014} (\bibinfo {year} {2024})}\BibitemShut {NoStop}%
\bibitem [{\citenamefont {Gamov}(1928)}]{gamov1928}%
  \BibitemOpen
  \bibfield  {author} {\bibinfo {author} {\bibfnamefont {G.~A.}\ \bibnamefont
  {Gamov}},\ }\bibfield  {title} {\bibinfo {title} {Quantum theory of the
  atomic nucleus},\ }\href@noop {} {\bibfield  {journal} {\bibinfo  {journal}
  {Zeitschrift fur Physik}\ }\textbf {\bibinfo {volume} {51}},\ \bibinfo
  {pages} {204} (\bibinfo {year} {1928})}\BibitemShut {NoStop}%
\bibitem [{\citenamefont {Hou}\ \emph {et~al.}(2024)\citenamefont {Hou},
  \citenamefont {Zhou}, \citenamefont {Xue}, \citenamefont {Yu}, \citenamefont
  {Han}, \citenamefont {Zhang},\ and\ \citenamefont {Lu}}]{Hou2024}%
  \BibitemOpen
  \bibfield  {author} {\bibinfo {author} {\bibfnamefont {Y.}~\bibnamefont
  {Hou}}, \bibinfo {author} {\bibfnamefont {J.}~\bibnamefont {Zhou}}, \bibinfo
  {author} {\bibfnamefont {M.}~\bibnamefont {Xue}}, \bibinfo {author}
  {\bibfnamefont {M.}~\bibnamefont {Yu}}, \bibinfo {author} {\bibfnamefont
  {Y.}~\bibnamefont {Han}}, \bibinfo {author} {\bibfnamefont {Z.}~\bibnamefont
  {Zhang}},\ and\ \bibinfo {author} {\bibfnamefont {Y.}~\bibnamefont {Lu}},\
  }\bibfield  {title} {\bibinfo {title} {Strain engineering of twisted bilayer
  graphene: The rise of strain‐twistronics},\ }\bibfield  {journal} {\bibinfo
   {journal} {Small}\ }\href {https://doi.org/10.1002/smll.202311185}
  {10.1002/smll.202311185} (\bibinfo {year} {2024})\BibitemShut {NoStop}%
\bibitem [{\citenamefont {Szyniszewski}\ \emph {et~al.}(2025)\citenamefont
  {Szyniszewski}, \citenamefont {Mostaani}, \citenamefont {Knothe},
  \citenamefont {Enaldiev}, \citenamefont {Ferrari}, \citenamefont {Fal’ko},\
  and\ \citenamefont {Drummond}}]{Szyniszewski2025}%
  \BibitemOpen
  \bibfield  {author} {\bibinfo {author} {\bibfnamefont {M.}~\bibnamefont
  {Szyniszewski}}, \bibinfo {author} {\bibfnamefont {E.}~\bibnamefont
  {Mostaani}}, \bibinfo {author} {\bibfnamefont {A.}~\bibnamefont {Knothe}},
  \bibinfo {author} {\bibfnamefont {V.}~\bibnamefont {Enaldiev}}, \bibinfo
  {author} {\bibfnamefont {A.~C.}\ \bibnamefont {Ferrari}}, \bibinfo {author}
  {\bibfnamefont {V.~I.}\ \bibnamefont {Fal’ko}},\ and\ \bibinfo {author}
  {\bibfnamefont {N.~D.}\ \bibnamefont {Drummond}},\ }\bibfield  {title}
  {\bibinfo {title} {Adhesion and reconstruction of graphene/hexagonal boron
  nitride heterostructures: A quantum monte carlo study},\ }\href
  {https://doi.org/10.1021/acsnano.4c10909} {\bibfield  {journal} {\bibinfo
  {journal} {ACS Nano}\ }\textbf {\bibinfo {volume} {19}},\ \bibinfo {pages}
  {6014} (\bibinfo {year} {2025})}\BibitemShut {NoStop}%
\end{thebibliography}%

\end{document}